%% file: cosmo12.tex
\documentclass[onecolumn,amsmath,amsssymb]{revtex4}
\input myepsf.tex

\def\gsim{ \lower .75ex \hbox{$\sim$} \llap{\raise .27ex \hbox{$>$}} }
\def\lsim{ \lower .75ex \hbox{$\sim$} \llap{\raise .27ex \hbox{$<$}} }

\begin{document}

\title{Detecting Dark Energy in Orbit -- the Cosmological Chameleon}

\author{Philippe Brax}
\affiliation{Service de Physique Th\'eorique, CEA-Saclay, Gif/Yvette cedex, France F-91191}

\author{Carsten van de Bruck}
\affiliation{Astro--Particle Theory and Cosmology Group, Department of Applied Mathematics,
University of Sheffield, Sheffield S3 7RH, UK}

\author{Anne-Christine Davis}
\affiliation{DAMTP, Centre for Mathematical Sciences,\\
Cambridge University, 
Wilberforce Road, Cambridge, CB3 OWA, UK}

\author{Justin Khoury and Amanda Weltman}
\affiliation{Institute for Strings, Cosmology, and Astroparticle Physics, 
Columbia University, New York, NY 10027 USA}

\begin{abstract}
We show that the chameleon scalar field can drive the current phase of cosmic acceleration for a large class of scalar potentials that are also consistent with local tests of gravity. These provide explicit realizations of a quintessence model where the quintessence scalar field couples directly to baryons and dark matter with gravitational strength. We analyze the cosmological evolution of the chameleon field and show the existence of an attractor solution with the chameleon following the minimum of its effective potential. For a wide range of initial conditions, spanning many orders of magnitude in initial chameleon energy density, the attractor is reached before nucleosynthesis. Surprisingly, the range of allowed initial conditions leading to a successful cosmology is wider than in normal quintessence. We discuss applications to the cyclic model of the universe and show how the chameleon mechanism weakens some of the constraints on cyclic potentials.
\end{abstract}

\maketitle

\section{Introduction} \label{intro}

A host of observations concord with the existence of a dark energy component with negative pressure, accounting for more than two thirds of the current energy budget. The evidence comes, for instance, from measurements of the cosmic microwave background temperature anisotropy~\cite{wmap} and Type Ia supernovae~\cite{sn1a}. While the data is so far consistent with the dark fluid being a cosmological constant, it is nevertheless interesting to consider the possibility that near future observations will reveal that $w$ differs from $-1$.

Having $w\neq -1$ implies that a parameter of the effective Lagrangian, namely the vacuum energy, is time-dependent. It follows from general covariance and locality that it must also be a function of space; in other words, the vacuum energy is a field, assumed for simplicity to be a fundamental scalar $\phi$. Scalar field models of dark energy generally come under the label of quintessence~\cite{quint}. Of course, this argument assumes that gravity is described by General Relativity (GR) for all relevant scales, and it is conceivable that the observed acceleration could result from a break down of GR on large scales~\cite{DGP,carroll,nima}. However, we focus on the former possibility.

Moreover, since $w\neq -1$ today, the vacuum energy must have varied significantly over the last Hubble time $H_0^{-1}$. This in turn requires $\phi$ to have a tiny mass of order $H_0\sim 10^{-33}$~eV. Indeed, if the mass is much smaller than $H_0$, then the field evolution is overdamped and the corresponding $w$ is unmeasurably close to -1; similarly, if the mass is much larger than $H_0$, the field is rolling too rapidly to cause cosmic acceleration. A natural question then arises: if such a nearly massless field exists, why have we not detected it in local tests of the Equivalence Principle (EP)~\cite{willbook} and fifth force searches~\cite{fischbach}? It is well-known that, in effective theories from string theory, such scalars generally couple directly to matter with gravitational strength, leading to unacceptably large violations of the EP.

Recently, two of us (JK and AW) have proposed a novel scenario~\cite{cham} which offers a natural resolution to this conflict. In this work they propose a scalar field which can evolve on a Hubble time today and cause cosmic acceleration, while coupling to matter with gravitational strength, in harmony with general expectations from string theory. The basic idea is that the scalar field acquires a mass which depends on the local background matter density. On Earth, where the density is high, the Compton wavelength of the field is sufficiently short to satisfy all existing tests of gravity; in the solar system, where the density is many orders of magnitude smaller, the Compton wavelength is larger than the size of the solar system; in the cosmos, where the density is tiny, the field can have a mass of order $H_0$ and cause cosmic acceleration. Because its physical properties depend sensitively on the environment, such a scalar field was dubbed chameleon. While the idea of a density-dependent mass term is not new~\cite{added,pol,others}, our work is novel in that the scalar field can couple directly to baryons with gravitational strength.

An important feature of the chameleon scenario is that it makes unambiguous and testable predictions for near-future tests of gravity in space. This is timely as three satellite experiments (SEE~\cite{SEE}, STEP~\cite{STEP} and GG~\cite{GG}) are in the proposal stage, while a fourth one (MICROSCOPE~\cite{MICRO}) will be launched in 2007. In the solar system, the chameleon is essentially a free field and thus mediates a long-range force. This force is very weak for large bodies, such as the Sun and the planets, therefore leaving planetary orbits nearly unperturbed. This is because of the thin-shell effect, detailed in~\cite{cham}. Intuitively, for sufficiently large objects, only a thin shell just beneath the surface contributes to the $\phi$-force on a test mass. This breakdown of the superposition principle is a consequence of the non-linear self-interactions of $\phi$. 

Typical test masses in the above satellite experiments, however, do not have a thin shell. Therefore, the extra force is comparable to their gravitational interaction. The chameleon model hence predicts that MICROSCOPE, STEP and GG could measure violations of the EP stronger than currently allowed by laboratory experiments. Furthermore, the SEE project could measure an effective Newton's constant different by order unity from that measured on Earth. Such outcomes would constitute strong evidence for the existence of chameleons in our Universe. Moreover, it is hoped that the real possibility of such surprising results will strengthen the scientific case for these missions.

In this paper, we discuss the cosmological history of a universe with a chameleon field. We prove the existence of an attractor solution, analogous to the tracker solution in quintessence models, which consists of the chameleon following the minimum of its effective potential. For a wide range of initial conditions, spanning many orders of magnitude in initial energy density for the scalar field, the solution converges to the attractor. While following the attractor, the energy density of the scalar field is always subdominant to the matter and radiation, except for when cosmic acceleration is triggered. The onset of the acceleration phase depends on the details of the potential.

We present a wide class of potentials for which acceleration occurs today; that is, for which the chameleon plays the role of quintessence. In doing so, we take advantage of the intriguing fact, showed in~\cite{cham}, that the largest value of $M$ allowed by existing tests of gravity is $10^{-3}$~eV, which coincides with the energy scale of dark energy. The constraint of $M\;\lsim\; 10^{-3}$ eV was derived for the chameleon in~\cite{cham} completely independently of any cosmological consideration. It is therefore remarkable and unexpected that the energy scale of dark energy emerges from a study of local tests of gravity.

Therefore, a natural class of scalar potentials for our purposes are of the form $V(\phi) = M^4f(\phi/M)$, with $M\approx 10^{-3}$~eV. That is, potentials that involve a single mass parameter $M$, which we tune to $10^{-3}$ eV, as required by observations. We stress that such a tuning is no better nor worse than the usual tuning of the cosmological constant or quintessence models.

\begin{figure}
\epsfxsize=3 in \centerline{\epsfbox{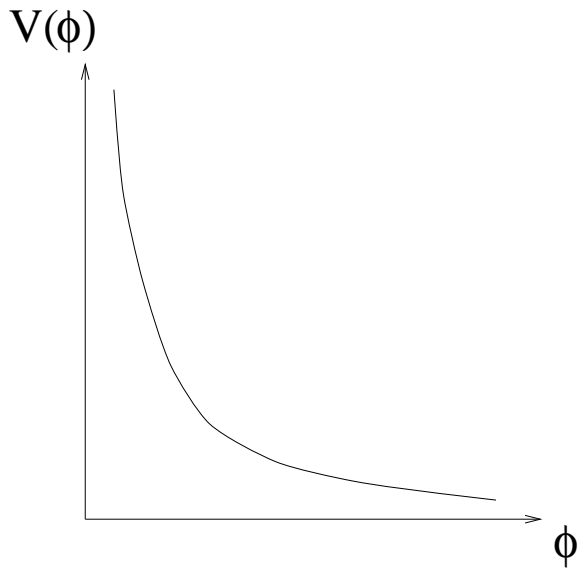}}
\caption{A typical scalar potential for the chameleon field.}
\label{fid}
\end{figure}

\begin{figure}
\epsfxsize=5 in \centerline{\epsfbox{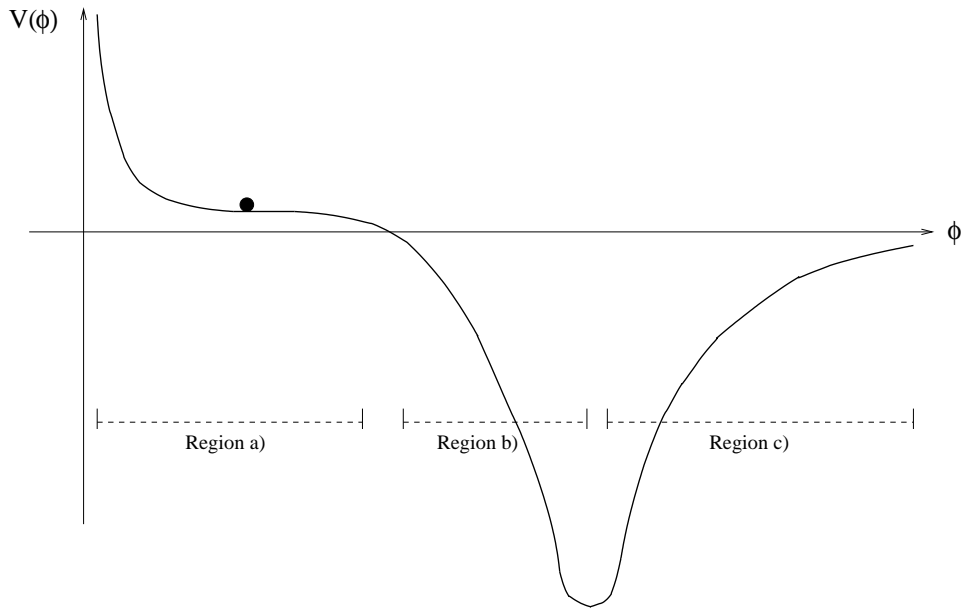}}
\caption{Scalar potential for cyclic models of the universe.
In this paper, we will only consider Region a).}
\label{cyclic}
\end{figure}

The function $f$ must satisfy only two broad requirements. Following~\cite{cham}, we assume that $(i)$ it is of the runaway form, and $(ii)$ it diverges at some finite value of $\phi$, which we take to be $\phi=0$ without loss of generality. (See~\cite{gubser} for an example of a successful chameleon model where neither of these conditions are satisfied.) Secondly, it must be flat and of order unity for today's value of the field, ensuring cosmic acceleration now. A fiducial example is $V(\phi) = M^4\exp(M^n/\phi^n)$, with $n$ some positive constant, which diverges at $\phi=0$ and tends to $M^4$ for $\phi\gg M$. See Fig.~\ref{fid}.
Notice that for pure power law potentials running away to zero at infinity, present-day cosmic acceleration with $M=10^{-3}$ eV cannot be obtained. Keep in mind, however, that it is not necessary for $V$ to tend to a constant as $\phi\rightarrow\infty$, as is the case for this particular choice of potential. For instance, the potential could become negative for larger field values, as illustrated in Fig.~\ref{cyclic}. Potentials of this form are of interest because of their direct applicability to cyclic models of the universe~\cite{cyclic,ek,seiberg,pert,design,rest}.

For potentials satisfying the above conditions, we show that the chameleon joins the attractor before the present epoch for a wide range of initial conditions. The resulting cosmology is then shown to be consistent with current observations. The most stringent constraint on initial conditions comes from the time-variation of coupling constants and masses since big bang nucleosynthesis (BBN).
Our analysis takes into account `kicks' due to species becoming 
non-relativistic in the radiation era. We find that the chameleon must be at
the minimum by the time of BBN or else the electron kick would induce an unacceptably large 
variation of particle masses. This requires that the chameleon varies by less than $0.1\; M_{\rm Pl}$ over the entire history of the universe. Translated in terms of the initial fractional energy density in the chameleon, $\Omega_\phi^{(i)}$, this requires $\Omega_\phi^{(i)}\;\lsim \; 0.1$, which is easily satisfied if one assumes equipartition at reheating. In particular, we see that this restriction on initial conditions is actually less restrictive than in normal quintessence~\cite{wang}.

While the cosmological evolution described below somewhat overlaps with previous studies of models of interacting dark energy and dark matter~\cite{added,dmde,amendola,moredmde} (henceforth DE-DM models), there are significant differences. The most important difference is that the chameleon not only couples to dark matter but also to baryons. Thus the chameleon model is subject to tight constraints from fifth force and EP experiments. These require that the mass scale $M$ in the potential be much smaller than generally considered in DE-DM models. This in turn results in weaker constraints from the cosmological evolution of the scalar field. Secondly, the coupling between the chameleon and matter fields is exponential (as in~\cite{amendola}) as opposed to linear (as in~\cite{dmde}). Thus, for small values of the chameleon field, this leads to weaker constraints from time-variation of masses and coupling. Thirdly, to our knowledge the discussion of the attractor solution and the approach to the attractor in terms of overshoot and undershoot solutions is new and parallels the corresponding treatment for quintessence~\cite{wang}. 

In Sec.~\ref{review}, we review the main ingredients of chameleon cosmology introduced in~\cite{cham} and give a brief review of how existing tests of gravity lead to the constraint $M\;\lsim\;10^{-3}$ eV. In Sec.~\ref{cos}, we focus on the cosmological evolution of the field for the above class of potentials and show the existence of an attractor solution. How the attractor solution is approached for general initial conditions is the subject of Sec.~\ref{over}; as in usual quintessence, we find two broad classes of solutions, so-called ``overshoot'' and ``undershoot'', corresponding respectively to whether the field begins to the left or to the right of the minimum of its effective potential.
We derive in Sec.~\ref{constraints} the range of initial conditions allowed by cosmological constraints and find that it spans many orders of magnitude in initial energy density of the chameleon. It turns out that the main constraint comes from the BBN bound on the time 
variation of particle masses. The behavior of the chameleon during inflation is considered in Sec.~\ref{inflation}. We investigate in Sec.~\ref{alpha} whether the chameleon can account for a time-varying fine-structure constant that has been suggested recently~\cite{webb}. In Sec.~\ref{cyclicpot} we apply our results to the cyclic model of the universe and show how the chameleon mechanism greatly expands the class of potentials suitable for cyclic cosmology. 

\section{Review of the chameleon model} \label{review}

The action governing the dynamics of the chameleon field $\phi$ is of the general form
\begin{equation}
S=\int d^4x\sqrt{-g}\left\{\frac{M_{\rm Pl}^2}{2}{\cal
R}-\frac{1}{2}(\partial\phi)^2- V(\phi)\right\}
- \int d^4x{\cal L}_{\rm m}(\psi_{\rm m}^{(i)},g_{\mu\nu}^{(i)})\,,
\label{action}
\end{equation}
where  $M_{\rm Pl}\equiv (8\pi G)^{-1/2}$ is the reduced Planck mass, $g$ is the determinant of the metric $g_{\mu\nu}$, ${\cal R}$ is the Ricci scalar and $\psi_{\rm m}^{(i)}$ are various matter fields labeled by $i$. A key ingredient of the model is the conformal coupling of $\phi$ with matter particles. More precisely, the excitations of each matter field $\psi_{\rm m}^{(i)}$ follow the geodesics of a metric $g_{\mu\nu}^{(i)}$ which is related to the Einstein-frame metric
$g_{\mu\nu}$ by the conformal rescaling
\begin{equation}
g_{\mu\nu}^{(i)}=e^{2\beta_i\phi/M_{\rm Pl}}g_{\mu\nu}\,,
\label{conformal}
\end{equation}
where $\beta_i$ are dimensionless constants~\cite{dgg}. In harmony with general expectations from string theory, we assume that the $\beta_i$'s are of order unity and different for each matter species. Varying the action with respect to $\phi$ yields the following Klein-Gordon equation
\begin{equation}
\nabla^2\phi = V_{,\phi} - \sum_i\frac{\beta_i}{M_{\rm Pl}}e^{4\beta_i\phi/M_{\rm Pl}}g_{(i)}^{\mu\nu}T^{(i)}_{\mu\nu}\,,
\label{eom0}
\end{equation}
where $T^{(i)}_{\mu\nu} = (2/\sqrt{-g^{(i)}})\delta {\cal L}_{\rm m}/\delta g_{(i)}^{\mu\nu}$ is the stress-energy tensor for the $i$th form of matter. 

For relativistic degrees of freedom, it is generally assumed that $T^{\mu}_{\mu} = 0$. This is not quite true however as the trace receives two corrections which play an important role in the evolution of the chameleon. First, in a cosmological context, each time a particle species becomes non-relativistic, the trace becomes significantly different from zero for about one e-fold of expansion~\cite{damnord,pol}. A second contribution comes from the trace anomaly~\cite{qcd,gravbar}. Until Sec.~\ref{over} we will ignore these two effects and neglect the relativistic fluid contribution to Eq.~(\ref{eom0}).

For non-relativistic matter with density $\tilde{\rho}_i$, one can make a perfect fluid approximation to obtain $g_{(i)}^{\mu\nu}T^{(i)}_{\mu\nu}\approx -\tilde{\rho}_i$. Defined in this way, however, $\tilde{\rho}_i$ is not conserved in Einstein frame. Instead, it is more convenient to define a matter density $\rho_i\equiv \tilde{\rho}_ie^{3\beta_i\phi/M_{\rm Pl}}$ which is independent of $\phi$ and conserved in Einstein frame. We therefore obtain
\begin{equation}
\nabla^2\phi = V_{,\phi} + \sum_i\frac{\beta_i}{M_{\rm Pl}}\rho_i e^{\beta_i\phi/M_{\rm Pl}}\,.
\label{eom}
\end{equation}

The key realization~\cite{cham,added,pol,others} from Eq.~(\ref{eom}) is that the dynamics of $\phi$ are not governed solely by $V(\phi)$, but rather by an effective potential:
\begin{equation}
V_{\rm eff}(\phi) = V(\phi) + \sum_i\rho_i e^{\beta_i\phi/M_{\rm Pl}} \,.
\label{veff}
\end{equation}
If $V$ is monotonically decreasing and $\beta_i > 0$ (or, equivalently, $V(\phi)$ monotonically increasing and $\beta_i<0$), this effective potential has a minimum at $\phi_{\rm min}$, satisfying
\begin{equation}
V_{,\phi}(\phi_{\rm min}) + \sum_i\frac{\beta_i}{M_{\rm Pl}}\rho_i e^{\beta_i\phi_{\rm min}/M_{\rm Pl}}=0\,.
\label{phimin}
\end{equation}
Meanwhile, the mass of small fluctuations about $\phi_{\rm min}$ is
\begin{equation}
m^2 = V_{,\phi\phi}^{\rm eff}(\phi_{\rm min}) = V_{,\phi\phi}(\phi_{\rm min}) + \sum_i\frac{\beta_i^2}{M_{\rm Pl}^2}\rho_i
e^{\beta_i\phi_{\rm min}/M_{\rm Pl}}\,.
\label{mmin}
\end{equation}
(See~\cite{sergio} for a stability analysis of the model.)

The self-interaction potential $V(\phi)$ is thought to arise from non-perturbative effects and is assumed to involve a single mass scale $M$:
\begin{equation}
V(\phi) = M^4f(\phi/M)\,,
\end{equation}
where $f$ is a dimensionless function. As in the original chameleon papers~\cite{cham}, we impose that the potential i) satisfy the tracker condition:
\begin{equation}
\Gamma\equiv \frac{V_{,\phi\phi}V}{V_{,\phi}^2} > 1\,,
\label{tracker}
\end{equation}
and ii) diverge at some finite value of $\phi$, denoted by $\phi=\phi_\star$. See Fig.~\ref{fid}. In Ref.~\cite{cham} it was believed that these were necessary for consistency with current tests of gravity. See~\cite{gubser}, however, for an example of a successful chameleon model where neither i) nor ii) holds. Thus the function $f$ satisfies
\begin{eqnarray}
\nonumber
& & \frac{f''f}{f'^2} > 1\,;\\
& & f \rightarrow \infty \qquad\;\;\; {\rm as}\;\;\;x\rightarrow x_\star \,.
\label{condsf}
\end{eqnarray}
For most of the paper, we shall assume $\phi_\star=0$ without loss of generality. For applications to cyclic cosmology in Sec.~\ref{cyclicpot}, however, we will need to choose $\phi_\star\gg M_{\rm Pl}$.

\begin{figure}
\epsfxsize=5 in \centerline{\epsfbox{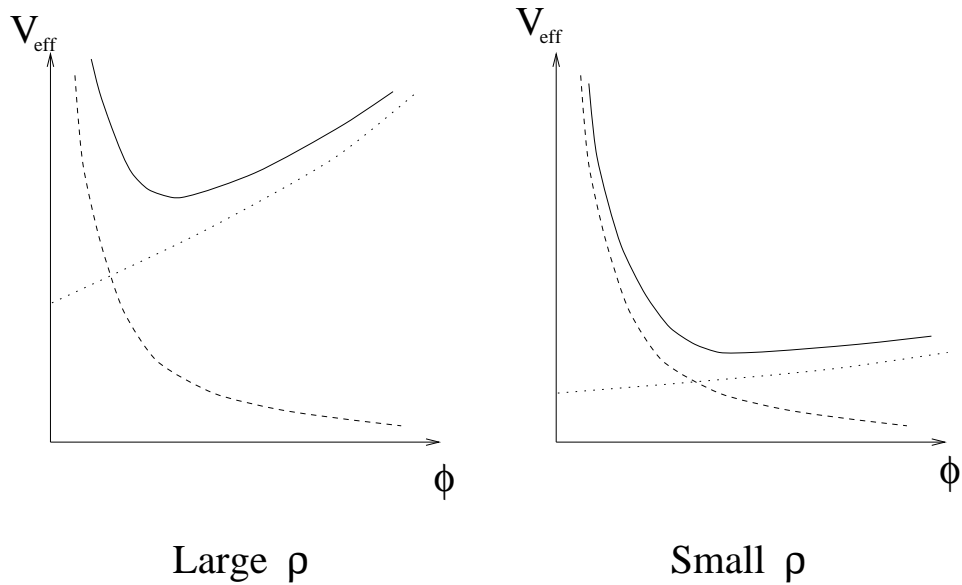}}
\caption{Chameleon effective potential for large and small matter density $\rho$. This illustrates that, as $\rho$ decreases, the minimum shifts to larger values of $\phi$ and the mass of small fluctuations decreases. The solid curve is the sum of the contribution from the actual potential $V(\phi)$ (dashed curve) and the contribution due to the matter density (dotted curve).}
\label{rhodep}
\end{figure}

An essential element of the model is the fact that $V_{\rm eff}$ depends explicitly on the matter density $\rho_i$, as seen in Eq.~(\ref{veff}). In particular, this implies that both $\phi_{\rm min}$ and $m$ are also functions of $\rho_i$. As illustrated in Fig.~\ref{rhodep}, for a general potential satisfying Eqs.~(\ref{condsf}), the mass is in fact an increasing function of $\phi$: the larger the density, the higher the mass. Thus, even though the chameleon mediates a composition-dependent fifth force of gravitational strength, it can satisfy the constraints from laboratory tests of the EP and fifth force by acquiring a sufficiently large mass locally. This is reviewed below.

\subsection{Constraint from fifth force and EP experiments} \label{Mrev}

The tightest constraint on the model comes from searches for a fifth force in the laboratory~\cite{fischbach}. The potential energy $U$ associated with fifth force interactions is generally parameterized by a Yukawa form:
\begin{equation}
U(r) =  - \alpha G M_1M_2\frac{e^{-r/\lambda}}{r}\,,
\label{fifthpot}
\end{equation}
where $M_1$ and $M_2$ are the masses of two test particles, $r$ is their
separation, $\alpha$ is the strength of the interaction, and $\lambda$ is the range.
Experiments have found no evidence for a fifth force for $\lambda\;\gsim 100\;\mu$m~\cite{adel}, assuming $\alpha\sim {\cal O}(1)$. Thus, in order for the chameleon model to be consistent with this result, we must impose that the range of the chameleon-interaction in the atmosphere, $m^{-1}_{\rm atm}$, be less than $100 \;\mu$m. Actually, this ignores the fact that fifth force experiments are performed in vacuum, where the density is much less than atmospheric density. Carefully taking into account the presence of the vacuum chamber actually results in a somewhat weaker bound of $m^{-1}_{\rm atm}\;\lsim\; 1$ mm. In~\cite{cham} it was shown that, for the inverse power-law potential $V(\phi) = M^{4+n}/(\phi-\phi_\star)^n$ with $n$ and $\beta$ of order unity, this constraint on $m_{\rm atm}$ translates into the following bound on $M$:
\begin{equation}
M\;\lsim\; 10^{-3}\;\;{\rm eV}\,.
\label{Mlim}
\end{equation}
See the Appendix for details. This condition not only ensures that the chameleon is consistent with fifth force searches, but it is also sufficient to satisfy all known local tests of GR, such as EP violation searches, Lunar Laser Ranging measurements, tests of post-Newtonian gravity and constraints on the spatial variation of coupling constants~\cite{cham}.

It is remarkable that the upper bound on $M$ in Eq.~(\ref{Mlim}) exactly coincides with the energy scale associated with the dark energy today. This comes as a complete surprise since the derivation of Eq.~(\ref{Mlim}) is based on local tests of gravity and is completely independent of cosmic acceleration. We will make use of this intriguing coincidence in turning the chameleon into a quintessence field driving the current phase of cosmic acceleration.

\subsection{Fiducial potential} \label{fidsec}

Unless stated otherwise, we henceforth take $\phi_\star = 0$. Since we
are interested in applications to quintessence, as our fiducial potential we choose
\begin{equation}
V(\phi) = M^4\exp(M^n/\phi^n)
\label{fidpot}
\end{equation}
with $M=10^{-3}$ eV. In the limit $\phi \;\gsim\; M$, this reduces to $V(\phi)\approx M^4 + M^{4+n}/\phi^n$. Since the constant term $M^4$ is only relevant dynamically on cosmological scales today, it can be dropped for the analysis of the tests of gravity in the laboratory and solar system. Thus Eq.~(\ref{Mlim}) derived in~\cite{cham} for the inverse power-law potential is virtually identical for our fiducial potential and hence satisfied for $M=10^{-3}$ eV.

It is illustrative to apply some of the general expressions above to our fiducial potential. For instance, since $V_{,\phi}=-n\phi^{-n-1}M^nV$ in this case, the field value at the minimum (Eq.~(\ref{phimin})) can be re-written in a form that will be useful for later applications
\begin{equation}
\left(\frac{M}{\phi_{\rm min}}\right)^{n+1} = \frac{\beta}{n}\frac{M}{M_{\rm Pl}}\frac{\rho_{\rm m}e^{\beta\phi_{\rm min}/M_{\rm Pl}}}{V(\phi_{\rm min})}\,,
\label{phimineg}
\end{equation}
where we have assumed a single matter component with density $\rho_{\rm m}$ and coupling $\beta$, for simplicity. Meanwhile, the mass of excitations about the minimum (Eq.~(\ref{mmin})) is given, once again for a single matter component, by
\begin{equation}
m^2 =  \frac{\beta \rho_{\rm m}e^{\beta\phi/M_{\rm Pl}}}{MM_{\rm Pl}} \left\{n\left(\frac{M}{\phi}\right)^{n+1}+(n+1)\frac{M}{\phi} +\beta \frac{M}{M_{\rm Pl}}\right\}\,,
\label{mtmp}
\end{equation}
where we have used Eq.~(\ref{phimin}).

The field value at the minimum today satisfies $\phi_{\rm min}^{(0)}\ll M_{\rm Pl}$. This is most easily seen by substituting $\rho_{\rm m}e^{\beta\phi_{\rm min}/M_{\rm Pl}}\sim V(\phi_{\rm min})\sim M^4$ in Eq.~(\ref{phimineg}). Assuming $\beta$ and $n$ are of order unity, we find
\begin{equation}
\phi_{\rm min}^{(0)}\sim \left(\frac{M}{M_{\rm Pl}}\right)^{n/(n+1)}M_{\rm Pl}\ll M_{\rm Pl}\,.
\label{phimin0}
\end{equation}
Furthermore, since $\phi_{\rm min}$ is an increasing function of time, the inequality $\phi_{\rm min}\ll M_{\rm Pl}$ holds for all relevant times, from the big bang until today. It follows that the mass in Eq.~(\ref{mtmp}) can be approximated by
\begin{equation}
m^2 \approx V_{,\phi\phi}(\phi_{\rm min}) = \frac{\beta \rho_{\rm m}e^{\beta\phi/M_{\rm Pl}}}{\phi M_{\rm Pl}} \left\{1+n+n\left(\frac{M}{\phi}\right)^{n}\right\}\,.
\label{mapprox}
\end{equation}

Finally, it will prove useful to determine the critical matter density $\rho_{\rm crit}$ for which $\phi_{\rm min}$ is of order $M$. That is, for $\rho\ll\rho_{\rm crit}$, one has $\phi_{\rm min}\gg M$, and the fiducial potential can be approximated as $V(\phi)\approx M^4 + M^{4+n}/\phi^n$. Substituting $\phi_{\rm min}\sim M\ll M_{\rm Pl}$ and $V(\phi_{\rm min})\sim M^4$ in Eq.~(\ref{phimineg}), and solving for $\rho_{\rm crit}$, we find

\begin{equation}
\rho_{\rm crit}\sim  \left(\frac{M_{\rm Pl}}{M}\right)\frac{nM^4}{\beta}\,.
\label{rhoc}
\end{equation}
For $\beta$ and $n$ of order unity and $M=10^{-3}$ eV, we find $\rho_{\rm crit} \approx 10^{-89}\;M_{\rm Pl}^4$, corresponding in cosmological terms to a temperature of $T_{\rm crit}\approx 10$~MeV and $\Omega_{\rm m}\approx 10^{-6}$, where $\Omega_{\rm m} \equiv \rho_{\rm m}\exp(\beta\phi/M_{\rm Pl})/3H^2M_{\rm Pl}^2$ is the fractional energy density in non-relativistic (matter) degrees of freedom.

\section{Cosmological evolution with a chameleon field} \label{cos}

In a homogeneous, isotropic and spatially flat universe, described by the Friedmann-Lema$\hat{\imath}$tre-Robertson-Walker metric $ds^2 = -dt^2 + a^2(t)d\vec{x}^2$, Eq.~(\ref{eom}) reduces to
\begin{equation}
\ddot{\phi} +3H\dot{\phi} =-V_{,\phi} - \frac{\beta}{M_{\rm Pl}}e^{\beta\phi/M_{\rm Pl}}\rho_{\rm m}\,,
\label{coseom}
\end{equation}
where dots represent derivatives with respect to cosmological time $t$.
For the moment we neglect the effect relativistic degrees of freedom
have on the trace of the stress tensor. This will be relaxed in Sec.~\ref{over}. The Hubble parameter $H$ is determined as usual by the Friedmann equation
\begin{equation}
3H^2M_{\rm Pl}^2 =\frac{1}{2}\dot{\phi^2} +V(\phi) + \rho_{\rm m}e^{\beta\phi/M_{\rm Pl}} + \rho_{\rm r} \,,
\end{equation}
where $\rho_{\rm m}$ and $\rho_{\rm r}$ denote respectively the energy density in matter and radiation. Both $\rho_{\rm m}$ and $\rho_{\rm r}$ are conserved in Einstein frame:
\begin{eqnarray}
\nonumber
& & \frac{\dot{\rho}_{\rm m}}{\rho_{\rm m}} = -3H  \\
& & \frac{\dot{\rho}_{\rm r}}{\rho_{\rm r}} = -4H \,.
\label{Econs}
\end{eqnarray}
Note that we have simplified the analysis by focusing on a single matter component. It is straightforward to allow for different matter species having different $\beta_i$, a generalization which does not substantially alter the results described below.

\subsection{Attractor solution} \label{attract}

We next show the existence of an attractor solution for the chameleon, consisting of the field following the minimum of the effective potential, $\phi=\phi_{\rm min}(t)$. Suppose that the field is initially at the minimum. An instant later, due to the redshifting of the matter density, the effective minimum will have moved to a slightly larger field value. Clearly, the characteristic time scale for this evolution is approximately a Hubble time, $H^{-1}$. Meanwhile, the response time for the field is given by $m^{-1}$, the period of oscillations about the minimum. From Eq.~(\ref{mapprox}), the ratio of these two time scales is then
\begin{equation}
\frac{m^2}{H^2} \approx 3\beta\Omega_{\rm m}\frac{M_{\rm Pl}}{\phi} \left\{1+n+n\left(\frac{M}{\phi}\right)^{n}\right\}\,.
\label{mfin}
\end{equation}

If $m\gg H$, then the response time, $m^{-1}$, of the field is much shorter than the characteristic time $H^{-1}$ over which the effective potential varies. In this case the chameleon adjusts itself and follows the minimum adiabatically as the latter evolves to larger field values. If, however, $m\ll H$, the response time is much larger than $H^{-1}$, and the field cannot follow the minimum. Instead, the chameleon starts to lag behind the minimum.

Below we show that, for $n$ and $\beta$ of order unity and $M=10^{-3}$ eV, we have $m\gg H$ from the big bang until today. Thus, if the field is initially at the minimum, it will follow the minimum as the latter evolves with time. Moreover, this solution is stable because if the field is slightly perturbed away from the minimum, it will oscillate and quickly settle back to the minimum. In other words, the solution $\phi=\phi_{\rm min}(t)$ is a dynamical attractor.

It is useful to compare this with usual quintessence, corresponding to setting $\beta=0$. Recall that quintessence also has an attractor called the {\it tracker solution}~\cite{wang}, $\phi=\phi_{\rm track}(t)$, which is defined by the condition
\begin{equation}
\frac{m^2}{H^2} \sim 1\,.
\end{equation}
That is, this corresponds to the field rolling down the potential at such a rate that its mass is always of order the Hubble constant. Our chameleon solution, $\phi=\phi_{\rm min}(t)$, however, satisfies $m\gg H$. Since $m$ is a monotonically decreasing function of $\phi$, it follows that
\begin{equation}
\phi_{\rm track}(t) > \phi_{\rm min}(t)
\end{equation}
at any given time $t$, {\it i.e.}, at any fixed value of $H$. In other words, if $\beta$ were zero (usual quintessence), the field would be driven to $\phi_{\rm track}$. For non-zero $\beta$, however, the dynamics of $\phi$ are governed by an effective potential with a minimum at $\phi_{\rm min}<\phi_{\rm track}$, thus preventing the field from reaching $\phi_{\rm track}$.

It remains to show that $m\gg H$ from the big bang (or inflationary reheating) until the present epoch. While this is easily done numerically, it is instructive to provide analytical arguments by studying Eq.~(\ref{mfin}) in two different regimes: $\phi\;\lsim\; M$ and $\phi\gg M$.

\begin{itemize}

\item $\phi\;\lsim\; M$: In this limit, Eq.~(\ref{mfin}) implies
\begin{equation}
\frac{m^2}{H^2} > 3\beta n \frac{M_{\rm Pl}}{M}\Omega_{\rm m}\,.
\label{mcase1}
\end{equation}
During the radiation and matter-dominated era, $\Omega_{\rm m}$ is a monotonically-increasing function of time. In particular, at the time when the universe has Planckian temperature (the worst case scenario for this argument), one has $\Omega_{\rm m}\approx 10^{-28}$. Thus, for all relevant times, we have $\Omega_{\rm m}\;\gsim \; 10^{-28}$. For $M=10^{-3}$ eV, Eq.~(\ref{mcase1}) therefore implies
\begin{equation}
\frac{m^2}{H^2} >  3\beta n \cdot 10^{2}\,,
\end{equation}
which is much larger than unity for reasonable values of $\beta$ and $n$.

\item $\phi\gg M$: In this regime, Eq.~(\ref{mfin}) reduces to
\begin{equation}
\frac{m^2}{H^2} \approx 3\beta (n+1) \frac{M}{\phi}\frac{M_{\rm Pl}}{M}\Omega_{\rm m}\,.
\label{mcase3}
\end{equation}
Moreover, since the potential in Eq.~(\ref{fidpot}) can be approximated as $V(\phi) \approx M^4$ in this limit, Eq.~(\ref{phimineg}) can be rewritten as
\begin{equation}
\left(\frac{M}{\phi}\right)^{n+1} \approx \frac{\beta}{n}\frac{M}{M_{\rm Pl}}\Omega_{\rm m}\frac{3H^2M_{\rm Pl}^2}{M^4}\,.
\end{equation}
Substituting this into Eq.~(\ref{mcase3}), we obtain
\begin{equation}
\frac{m^2}{H^2} \sim\left\{\left(\frac{M_{\rm Pl}}{M}\right)^n\Omega_{\rm m}^{n+2}\frac{3H^2M_{\rm Pl}^2}{M^4}\right\}^{\frac{1}{n+1}}\,,
\label{forlater}
\end{equation}
where we have neglected a prefactor of order unity. Now $3H^2M_{\rm Pl}^2$ is an increasing function of redshift and, given our choice of $M=10^{-3}$ eV, is of order $M^4$ today. Hence, from the big bang until today,
\begin{equation}
\frac{m^2}{H^2}\;\gsim \;\left\{\left(\frac{M_{\rm Pl}}{M}\right)^n\Omega_{\rm m}^{n+2}\right\}^{\frac{1}{n+1}}\,.
\end{equation}
Recall from the end of Sec.~\ref{fidsec} that $\Omega_{\rm m}\sim 10^{-6}$ when $\phi_{\rm min}\sim M$. Since $\Omega_{\rm m}$ is an increasing function of time in the radiation and matter-dominated era, it follows that $\Omega_{\rm m}\gg 10^{-6}$ in the regime $\phi\gg M$. Moreover, substituting $M=10^{-3}\;{\rm eV}\approx 10^{-30}M_{\rm Pl}$, we obtain
\begin{equation}
\frac{m^2}{H^2} \gg  10^{12(2n-1)/(n+1)}\,,
\end{equation}
which is greater than unity for $n\;\gsim\; {\cal O}(1)$. This proves that $m > H$ in the regime $\phi\gg M$ as well.

\end{itemize}

Thus we have shown that $\phi=\phi_{\rm min}(t)$ is an attractor from the big bang until today. What will happen to the chameleon in the future? Currently, since $m\gg H$, the field is still at the minimum of the effective potential, while the universe undergoes cosmic acceleration. Very soon, however,the energy density becomes completely dominated by the vacuum energy ($V\approx M^4$), and the cosmic evolution is driven towards de Sitter. In the process, the matter energy density redshifts away at an exponential rate. Eventually, it is so dilute that one reaches a point where $m\sim H$, as seen from Eq.~(\ref{mfin}). When this happens, the field can no longer follow the minimum. The dynamics of $\phi$ are then determined to a good approximation solely by its potential, $V(\phi)$, and the evolution of the chameleon converges to that of normal quintessence.

\subsection{Dynamics of $\phi$ along the attractor}

Having shown that $\phi=\phi_{\rm min}(t)$ is a dynamical attractor until the present epoch, we next would like to argue that the field is slow-rolling as it follows the attractor and derive an expression for its equation of state.

It is straightforward to show that $\dot{\phi}^2/2\ll V$ as the chameleon follows the minimum. Recall from Eq.~(\ref{phimin}) that $\phi_{\rm min}$ satisfies $-V_{\phi}(\phi_{\rm min}) = \beta\rho_{\rm m}\exp(\beta\phi_{\rm min}/M_{\rm Pl})/M_{\rm Pl}$. Taking time derivatives on both sides and using the fact, shown in Eq.~(\ref{mapprox}), that $m^2\approx V_{,\phi\phi}\gg -V_{,\phi}\beta/M_{\rm Pl}$, we find
\begin{equation}
\dot{\phi}_{\rm min} \approx -\frac{3HV_{,\phi}}{V_{,\phi\phi}}\,.
\label{dotphimin}
\end{equation}
It therefore follows that
\begin{equation}
\frac{\dot{\phi}_{\rm min}^2}{2V(\phi_{\rm min})}\approx \frac{9}{2}\frac{H^2}{V_{,\phi\phi}}\left(\frac{V_{,\phi}^2}{V_{,\phi\phi}V}\right)\approx \frac{9}{2}\frac{H^2}{m^2}\frac{1}{\Gamma}\,,
\label{kinpot}
\end{equation}
where in the last step we have once again used Eq.~(\ref{mapprox}) and substituted $\Gamma\equiv V_{,\phi\phi}V/V_{,\phi}^2$ (see Eq.~(\ref{tracker})). If we recall from Sec.~\ref{attract} that $m^2\gg H^2$ and from Eq.~(\ref{tracker}) that $\Gamma > 1$, we obtain
\begin{equation}
\frac{\dot{\phi}_{\rm min}^2}{2V(\phi_{\rm min})}\ll 1\,,
\label{slowroll}
\end{equation}
which proves that the field is slow-rolling along the attractor.

Next we derive an expression for the equation of state of $\phi$ as it follows the minimum. Since $\phi$ is a non-minimally coupled scalar, however, this is not given by the usual expression $w= (\dot{\phi}^2 -2V)/(\dot{\phi}^2+2V)$. Instead, we must compute $w$ directly from the time evolution of $\rho_\phi\equiv \dot{\phi}^2/2+V$, the energy density in $\phi$. Equation~(\ref{slowroll}) implies that $\rho_\phi\approx V$ at the minimum, and therefore
\begin{equation}
\frac{\dot{\rho}_\phi}{\rho_\phi}\approx \frac{V_{,\phi}}{V}\dot{\phi}_{\rm min} = -3H\Gamma^{-1}\,,
\label{dotrho}
\end{equation}
where in the last step we have substituted Eqs.~(\ref{tracker}) and~(\ref{dotphimin}). In analogy with the usual energy conservation equation, we can define an effective equation of state, $w_{\rm eff}$, for $\phi$ as
\begin{equation}
\frac{\dot{\rho}_\phi}{\rho_\phi} \equiv -3H(1+w_{\rm eff})\,,
\end{equation}
from which we can read off
\begin{equation}
w_{\rm eff} = -1 + \Gamma^{-1}\,.
\label{weff}
\end{equation}

Equation~(\ref{weff}) is the main result of this Section. It holds for a large class of runaway potentials satisfying Eqs.~(\ref{condsf}). Indeed, the only extra assumption in deriving this result is that $\phi_{\rm min}\ll M_{\rm Pl}$ for all relevant times, which implies $m^2\approx V_{,\phi\phi}(\phi_{\rm min})$.

Applying this result to our fiducial potential, $V(\phi) = M^4\exp(M^n/\phi^n)$, only a few steps of algebra are necessary to show that
\begin{equation}
\Gamma(\phi_{\rm min}) = 1 + \left(1+\frac{1}{n}\right)\left(\frac{\phi_{\rm min}}{M}\right)^n\,.
\end{equation}
Substituting this in Eq.~(\ref{weff}), we find that $w_{\rm eff}\approx 0$ for $\phi_{\rm min}\ll M$, while $w_{\rm eff}\approx -1$ for $\phi_{\rm min}\gg M$. Recall from the discussion at the end of Sec.~\ref{fidsec} that $\phi_{\rm min}\sim M$ when the universe has temperature of order $T\sim 10$ MeV. Thus, for this choice of potential, we have shown that the chameleon behaves like dust ($w_{\rm eff}\approx 0$) for $T\;\gsim\; 10$ MeV, and like vacuum energy ($w_{\rm eff}\approx -1$) for $T\;\lsim\; 10$ MeV.

Coming back to the case of general $V(\phi)$, it is instructive to compare Eq.~(\ref{weff}) with the usual expression for the equation of state, $w_{\rm usual}= (\dot{\phi}^2 -2V)/(\dot{\phi}^2+2V)$. From Eq.~(\ref{kinpot}), we obtain
\begin{equation}
w_{\rm usual}\approx -1+  \frac{\dot{\phi}^2}{V}\approx -1 + 9\frac{H^2}{m^2}\frac{1}{\Gamma}\,.
\end{equation}
Since the second term is down by a factor of $H^2/m^2\ll 1$ compared to its counterpart in Eq.~(\ref{weff}), it follows that $w_{\rm usual}$ is much closer to $-1$ than $w_{\rm eff}$. But, as argued above, it is really $w_{\rm eff}$ that controls the time-evolution of the chameleon's energy density.

We can also compare Eq.~(\ref{weff}) with the corresponding expression for the equation of state of normal quintessence. Along the tracker solution, the equation of state $w_{\rm Q}$ of the quintessence field is~\cite{wang}
\begin{equation}
w_{\rm Q}\approx \frac{w_{\rm B} -2(\Gamma-1)}{1+2(\Gamma-1)}\,,
\label{wQ}
\end{equation}
where $w_{\rm B}$ is the equation for the background perfect fluid ({\it i.e.}, $w_{\rm B}\approx 1/3$ during the radiation-dominated era; $w_{\rm B}\approx 0$ during the matter-dominated era). In contrast with $w_{\rm Q}$, the chameleon's effective equation of state is independent of $w_{\rm B}$. This is because the location of the minimum in chameleon cosmology is nearly independent of the energy density in relativistic degrees of freedom, since these have a nearly traceless stress-energy tensor and therefore contribute negligibly to the right-hand side of Eq.~(\ref{eom0}). Consequently, the evolution of the chameleon along the attractor is insensitive to whether or not the matter density dominates over radiation.

\section{Approaching the attractor} \label{over}

In this Section we describe how the attractor solution is approached for general initial conditions. For simplicity we assume the kinetic energy of the field is initially zero, although it is trivial to generalize the discussion to the case of non-zero kinetic energy. Moreover, we focus on the case where the energy density in $\phi$ is initially less than the energy density in matter and radiation, as expected from equipartition at reheating. More pragmatically, however, we will see in Sec.~\ref{constraints} that this is required in order for the cosmology to be consistent with the measured abundance of light elements.

So let us start the field from some arbitrary value $\phi_i$ at time $t_i$. The effective potential at that time displays a minimum at $\phi_{\rm min}(t_i)$, with $\phi_{\rm min}$ determined by Eq.~(\ref{phimin}). If the field is released at the minimum, $\phi_i = \phi_{\rm min}(t_i)$, then the field just keeps following the minimum until today, as argued in Sec.~\ref{attract}. Similarly, if the field starts very near $\phi_{\rm min}(t_i)$, it oscillates and quickly settles to the minimum.

More generally we wish to consider the limiting cases where $\phi_i\ll \phi_{\rm min}(t_i)$ and $\phi_i\gg \phi_{\rm min}(t_i)$. As it will soon become clear, and to make a parallel with the analogous discussion in usual quintessence, the corresponding solutions will be referred to as {\it overshoot} and {\it undershoot}, respectively.

\subsection {Undershooting} \label{undershoot}

In this case $\phi_i\gg \phi_{\rm min}(t_i)$, the $V_{,\phi}$ source term can be neglected, and the equation for $\phi$ reduces to
\begin{equation}
\ddot{\phi} + 3H\dot{\phi}  \approx \frac{\beta}{M_{\rm Pl}} T^{\mu}_{\mu}\,,
\label{frozen}
\end{equation}
where we have reintroduced the trace of the stress tensor. If the trace were negligibly small for relativistic degrees of freedom, then one would have $T^{\mu}_{\mu} \approx -\rho_{\rm m}$, where $\rho_{\rm m}$ is the non-relativistic matter density. But then, during the radiation-dominated era, the latter would be utterly negligible compared to the friction term, $3H\dot{\phi}$. Thus the field would be overdamped and would remain essentially frozen at its initial value $\phi_i$.

Fortunately, however, the trace is not always small for a realistic relativistic fluid. As the universe expands and cools, each massive particles species successively becomes non-relativistic whenever $m\sim T$. When this happens, the trace becomes non-zero for about one e-fold of expansion, thus driving the field a bit closer to the minimum~\cite{damnord}. To see this explicity, we follow~\cite{damnord,pol} and note that each component, labeled by $i$, of the relativistic plasma  contributes
\begin{equation}
T^{\mu\;(i)}_{\mu} = -\frac{45}{\pi^4}H^2M_{\rm Pl}^2\frac{g_i}{g_\star(T)}\tau(m_i/T)\,,
\label{Tdam}
\end{equation}
where $g_\star(T) = \sum_{\rm bosons} g_i^{\rm boson}(T_i/T)^4 + (7/8)\sum_{\rm fermions}g_i^{\rm fermion}(T_i/T)^4$ is the usual expression for the effective number of relativistic degrees of freedom, while $g_i$ and $T_i$ are the number of degrees of freedom and temperature of the $i^{\rm th}$ species, respectively. The function $\tau$ is defined by
\begin{equation}
\tau(x) = x^2\int_x^{\infty} du\frac{\sqrt{u^2-x^2}}{e^u \pm 1}\,,
\end{equation}
where the $\pm$ sign is for fermions and bosons, respectively. It is negligibly small both for $x\ll 1$ and $x\gg 1$, but becomes of order unity when $x\sim {\cal O}(1)$.

Integrating Eq.~(\ref{frozen}) numerically, one finds that the total displacement in $\phi$ due to the source term in Eq.~(\ref{Tdam}) is approximately given by
\begin{equation}
(\Delta\phi)_i \approx -\beta \frac{g_i}{g_\star(m_i)}
\left(
\begin{matrix}
7/8 \cr
1
\end{matrix}
\right)M_{\rm Pl}\,,
\label{Deli}
\end{equation}
where the upper and lower numerical coefficients are for fermions and bosons respectively, and $g_\star(m_i)$ is the effective number of relativistic degrees of freedom when $T=m_i$. This can also be derived analytically by approximating $\tau(x)$ as a $\delta$-function source in Eq.~(\ref{Tdam}), as shown in the Appendix (Sec.~\ref{apkick}). Thus, each species, whenever it becomes non-relativistic, effectively gives a ``kick'' to the chameleon, driving the latter closer to the minimum of the effective potential.

\begin{table*}[htb]
\small
\hbox to \hsize{\hfil\begin{tabular}{|c|c|}
\hline
\hspace{5pt}Particle	& $g_i/g_\star$ \\
\hline
\hspace{5pt} $t$	& $12/106.75$\\
\hspace{5pt} $Z$	& $3/95.25$\\
\hspace{5pt} $W^{\pm}$	& $6/92.25$\\
\hspace{5pt} $b$	& $12/86.25$\\
\hspace{5pt} $\tau$	& $4/75.75$\\
\hspace{5pt} $c$	& $12/72.25$\\
\hspace{5pt} $\pi$	& $3/17.25$\\
\hspace{5pt} $\mu$	& $4/14.25$\\
\hspace{5pt} $e$	& $4/10.75$\\
\hline
\end{tabular}\hfil}
\caption{List of particle species, in order of decreasing mass threshold, that contribute to kicking the chameleon towards the minimum.}
\label{partg}
\end{table*}

We can calculate the maximal displacement, $(\Delta\phi)_{\rm tot}$, by summing over all relevant massive particle species. In doing so, we only include particles that become non-relativistic at a much lower temperature scale than that of the phase transition through which they acquired a mass. See Table~\ref{partg} for a list of the relevant particles and corresponding $g_i$'s. We will see below that big bang nucleosynthesis (BBN) constrains the chameleon to be at the minimum by the onset of nucleosynthesis. This will be the case if $(\phi_i -\phi_{\rm min}^{({\rm BBN})}) \;\lsim\; |(\Delta\phi)_{\rm tot}|$, where $\phi_{\rm min}^{({\rm BBN})}$ is the location of the minimum at BBN. Hence, in doing the sum, we must neglect the electron contribution since its mass threshold is too close to the temperature during BBN ($\sim 1$~MeV). With this {\it proviso}, we can substitute the results of Table~\ref{partg} in Eq.~(\ref{Deli}) and obtain
\begin{equation}
(\Delta\phi)_{\rm tot} \approx - \beta M_{\rm Pl}\,.
\label{Delphitot}
\end{equation}
Thus, for $\beta\sim {\cal O}(1)$, the cumulative effect of the kicks is to push the field a distance of order $M_{\rm Pl}$ towards the minimum.

\begin{figure}
\epsfxsize=5 in \epsfysize=5 in \centerline{\epsfbox{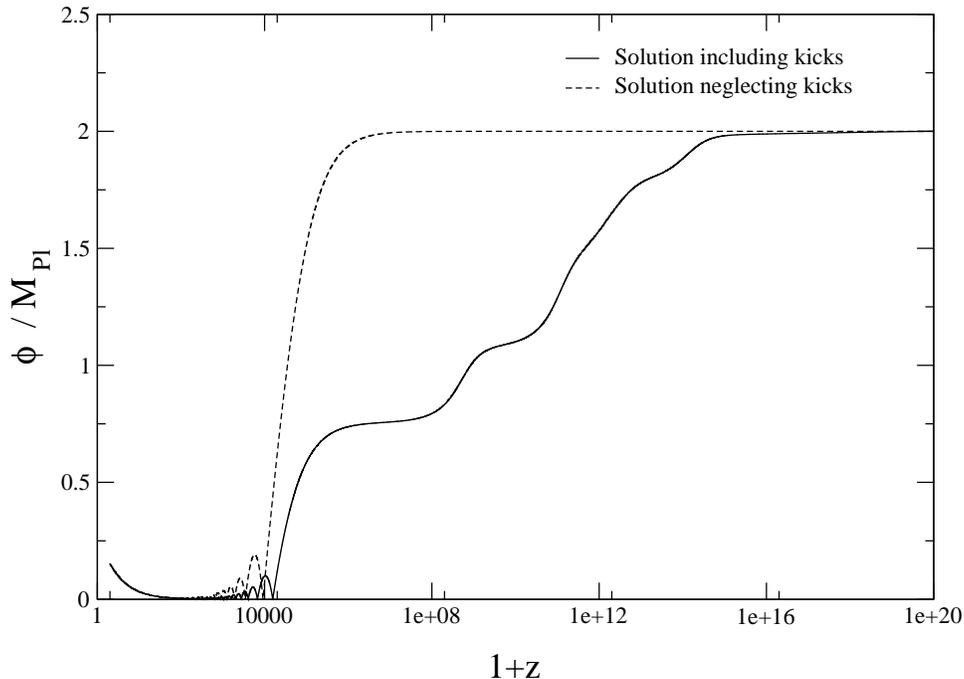}}
\caption{Evolution of the chameleon field $\phi$ as a function of $1+z$. In this example the 
potential is $V(\phi) = M^6/\phi^2$, $\beta = 1$ and $M = 10$ MeV, the latter being larger than the required $M=10^{-3}$~eV due
to numerical limitations. The initial conditions for the field are $\phi_i = 2 \;M_{\rm Pl}$ and 
$\dot \phi_i = 0$ at $z=10^{20}$. The dashed line presents the case in which the contribution of particles which become 
non-relativistic is neglected. The solid line includes the contribution of the particles in Table \ref{partg} to $T^{\mu}_\mu$. Since $\phi_i \;\gsim\;\beta M_{\rm Pl}$ in this case, the field does not reach the minimum by the onset of BBN ($z\approx 10^9$). As explained in the text, this solution is therefore ruled out by BBN constraints on time-variation of particle masses (see Sec.~\ref{constraints}).}
\label{carsten}
\end{figure}

The results of numerical integration are shown in Fig.~\ref{carsten} for $\phi_i=2\;M_{\rm Pl}$, where we compare the solution with and without the kicks described above. Because of numerical limitations, we were restricted to choosing $\phi_i\;\gsim\; M_{\rm Pl}$ only. For the solution without kicks, the chameleon remains frozen at its initial value throughout the radiation-dominated epoch. For the solution including kicks, however, it is pushed to smaller field values as various particle species successively become non-relativistic. It is seen that the total displacement from $z=10^{20}$ to $z=10^9$ (redshift of BBN) is indeed of order $\sim M_{\rm Pl}$, consistent with Eq.~(\ref{Delphitot}).

As we will see later, the particular solution shown in Fig.~\ref{carsten} is ruled out by BBN constraints since the field is not at the minimum by the onset of BBN. This is because numerical limitations forced us to choose $\phi_i\;\gsim\; M_{\rm Pl}$, a range of initial values larger than the total displacement given in Eq.~(\ref{Delphitot}). Thus, in this case, the field begins so far away from the minimum that the sequence of kicks is insufficient to bring it in the vicinity of the minimum by the onset of BBN. Even though this particular solution is ruled out, a few instructive comments can be made. After the electron has become non-relativistic ($z\approx 10^9$), the chameleon receives no further ``kicks'' and thus is essentially stuck at some particular value. It remains frozen there until matter-radiation equality ($z\approx 10^4$). At that time, one has $H^2M_{\rm Pl}^2\sim \rho_{\rm m}\approx T^{\mu}_{\mu}$, and the driving term in Eq.~(\ref{frozen}) is then comparable in magnitude with the friction term. Thus the field begins to roll towards the minimum, undergoes large anharmonic oscillations, and eventually converges to the minimum. These oscillations will be described in more detail in Sec.~\ref{conv}.

Had the numerics allowed us to probe the regime $\phi_i\;\lsim\; M_{\rm Pl}$, the solution would initially look similar to that of Fig.~\ref{carsten}, with each kick bringing the field closer to the minimum. Eventually, however, one of the kicks pushes the field sufficiently close to the minimum. The chameleon then starts oscillating and quickly settles to the minimum. Subsequent kicks generate oscillations about the minimum which are rapidly damped by the expansion of the universe. In particular, for the case of the electron contribution which kicks in during BBN, these oscillations are sufficiently small in amplitude to obey BBN constraints on time-variation of particle masses. This is discussed in more detail in the Appendix.

To summarize, if $\phi_i\;\gsim\;\beta M_{\rm Pl}$, the chameleon is initially so far from the minimum of the effective potential that the sequence of kicks is insufficient to push it in the vicinity of the minimum. In particular, the field is still away from the minimum by the onset of BBN, and, as we will see in Sec.~\ref{constraints}, this solution is ruled out by BBN constraints on time-variation of particle masses. If $\phi_i\;\lsim\;\beta M_{\rm Pl}$, then the kicks eventually push the chameleon near the minimum. At that point, the field oscillates and quickly settles down to the minimum. In particular, it is at the minimum by the onset of BBN, and this solution is consistent with BBN constraints, as explained in Sec.~\ref{constraints}.

We conclude by noting that another potentially important contribution to $T^{\mu}_{\mu}$ during the radiation-dominated era is from the trace anomaly. It was shown in~\cite{qcd} (see also~\cite{gravbar}) that the effective equation of state for a plasma of an SU($N_c$) with coupling $g$ and $N_f$ flavors is given by
\begin{equation}
1-3w = \frac{5}{6\pi^2}\left(\frac{g^2}{4\pi}\right)^2 \frac{(N_c + \frac{5}{4}N_f)(\frac{11}{3}N_c-\frac{2}{3}N_f)}{2+\frac{7}{2}N_cN_f/(N_c^2-1)} + {\cal O}(g^5)\,.
\end{equation}
For QCD with $N_c=3$ and $N_f = 6$, the above is of order $10^{-3}$ for energies above 100 GeV ({\it i.e.,} in the perturbative regime of QCD). This is too small to yield a significant displacement for the chameleon. However, larger gauge groups and $N_f$ could result in an important driving term near the unification scale~\cite{gravbar}, thereby improving on Eq.~(\ref{Delphitot}). For the purpose of this paper, however, we take a conservative approach and restrict ourselves to Standard Model degrees of freedom.

\subsection {Overshooting} \label{overshoot}

Since $\phi_i\ll \phi_{\rm min}(t_i)$, in this case it is the $\rho_{\rm m}$ source term that is negligible in Eq.~(\ref{coseom}). Thus the equation for $\phi$ can be approximated by
\begin{equation}
\ddot{\phi} + 3H\dot{\phi}  \approx - V_{,\phi}\,,
\end{equation}
which describes the evolution of a minimally-coupled scalar field. Once again, the dynamics are governed by a friction force, $3H\dot{\phi}$, and a driving force, $-V_{,\phi}$. However, the driving term initially dominates over the friction term in this case. To see this, note that $\phi_i\ll \phi_{\rm min}(t_i)$ implies $V_{,\phi\phi}(\phi_i)\gg m^2_i$, where $m^2_i$ is the mass of small fluctuations about $\phi_{\rm min}(t_i)$. Moreover, we have argued in Sec.~\ref{attract} that $m^2\gg H^2$ for all relevant times, therefore $V_{,\phi\phi}(\phi_i)\gg H^2_i$, where $H_i$ is the initial value of the Hubble parameter. In other words, the field is underdamped, and its evolution is essentially that of a free field. 

Thus very quickly its energy becomes kinetic-dominated, $\dot{\phi}^2\gg V$. Since $\phi_i$ is much smaller than $\phi_{\rm min}$, the field then rolls past the minimum and thus {\it overshoots}. It keeps on rolling until its kinetic energy has sufficiently redshifted so that the Hubble damping term becomes important. Then the  field essentially comes to a halt at some value $\phi_{\rm stop} > \phi_{\rm min}$, which we can estimate as follows. Since the energy in $\phi$ is kinetic-dominated as it rolls, we have $\dot{\phi}\sim a^{-3}$. Moreover, for the range of initial conditions relevant to this discussion, we may assume the universe is radiation-dominated, and therefore $a\sim t^{1/2}$. Combining these two facts, a little algebra shows that
\begin{equation}
a\frac{d\phi}{da}  = \sqrt{6\Omega_{\phi}^{(i)}}\left(\frac{a_i}{a}\right) M_{\rm Pl}\,,
\end{equation}
where $\Omega_{\phi}^{(i)}$ and $a_i$ are the initial fractional energy density and initial value of the scale factor, respectively. With initial condition $\phi=\phi_i$ at $a=a_i$, this integrates to
\begin{equation}
\phi(a) = \phi_i + \sqrt{6\Omega_{\phi}^{(i)}}M_{\rm Pl}\left(1-\frac{a_i}{a}\right) \,,
\end{equation}
which gives $\phi\rightarrow\phi_{\rm stop}$ in the limit $a\gg a_i$, where
\begin{equation}
\phi_{\rm stop} \approx \phi_i + \sqrt{6\Omega_{\phi}^{(i)}}M_{\rm Pl}\,.
\label{phistop}
\end{equation}
This is precisely what one finds for the overshoot solution in usual quintessence ({\it e.g.}, see Eq.~(11) of~\cite{wang}), which is not surprising since the non-minimal coupling of the chameleon is irrelevant in the above derivation. Once the chameleon reaches $\phi_{\rm stop}$, the solution is then exactly as in the undershoot case above, with $\phi_i$ replaced by $\phi_{\rm stop}$. 

\subsection{Converging to the minimum}\label{conv}

When the field reaches the vicinity of the minimum, it begins oscillating and eventually converges to the minimum. In this Section, we study these oscillations and the rate of convergence. To proceed analytically, we assume that the linear approximation is valid so the oscillations are harmonic. Strictly speaking, this of course only holds when the field is sufficiently close to the minimum. Nevertheless, we will find numerically that the range of validity is actually much wider and applies even when the oscillations are apparently anharmonic. Moreover, we neglect the kinetic energy from the last ``kick'' that drove the field in the vicinity of the minimum.

In the linear approximation, the effective potential is given by
\begin{equation}
V_{\rm eff}\approx \frac{1}{2}m^2(t)(\phi(t)-\phi_{\rm min})^2\,,
\end{equation}
where we have dropped the vacuum energy term, which is completely irrelevant for the present discussion. Meanwhile the total energy density in $\phi$ is well approximated by
\begin{equation}
\rho_\phi\approx \frac{1}{2}(\dot{\phi}-\dot{\phi}_{\rm min})^2 + \frac{1}{2}m^2(t)(\phi(t)-\phi_{\rm min})^2\,.
\end{equation}
Note that both $\phi_{\rm min}$ and $m$ are functions of time, a fact that will be 
important in the discussion below. Since the motion is assumed harmonic, averaging over a few oscillations gives
\begin{equation}
\rho_\phi\approx \frac{1}{2}m^2(t)<(\phi(t)-\phi_{\rm min})^2>\,.
\label{rhoavg}
\end{equation}

Since $m\gg H$, the problem is analogous to a pendulum which is slowly lengthened~\cite{birrell}. It is well known that in this adiabatic approximation the total oscillator number,
\begin{equation}
{\cal N} \equiv \frac{\rho_\phi a^3}{m(t)}\,,
\end{equation}
is conserved. Substituting Eq.~(\ref{rhoavg}), it follows that
\begin{equation}
m(t)<(\phi(t)-\phi_{\rm min})^2> \sim a^{-3}\,.
\label{interm}
\end{equation}
See~\cite{wilczek} and the Appendix for alternative derivations. 

If $m$ were constant, Eq.~(\ref{interm}) would imply that the energy density in the oscillations redshifts like dust, which is not surprising. In the chameleon model, however, this quantity is time-dependent. Recall from Eq.~(\ref{mapprox}) that $m\approx V_{,\phi\phi}^{1/2}$, and thus
\begin{equation}
\frac{\dot{m}}{m}\approx \frac{1}{2}\frac{V_{,\phi\phi\phi}}{V_{,\phi\phi}}\dot{\phi}_{\rm min}\approx -\frac{3H}{2}\frac{V_{,\phi\phi\phi}V_{,\phi}}{V_{,\phi\phi}^2}\,,
\end{equation}
where in the last step we have substituted Eq.~(\ref{dotphimin}). For the inverse power-law potential, $V\sim \phi^{-n}$, this reduces to
\begin{equation}
\frac{\dot{m}}{m}\approx  -\frac{3H}{2}\frac{n+2}{n+1}\,,
\label{mdot}
\end{equation}
from which we conclude that
\begin{equation}
m\sim a^{-3(n+2)/2(n+1)}\,.
\label{mbehave}
\end{equation}
Substituting this in Eq.~(\ref{interm}), we obtain
\begin{equation}
<(\phi-\phi_{\rm min}(t))^2>^{1/2} \sim a^{-3n/4(n+1)}\,.
\label{tocompare}
\end{equation}
It is worth mentioning that the time-dependence of $<(\phi-\phi_{\rm min})^2>$ is independent of $M$. The parameter $M$ only enters implicitly in that $\phi_{\rm min}$ at any fixed time depends on the choice of $M$. Thus, for fixed $\phi_i$, the difference $<(\phi-\phi_{\rm min})^2>$ is initially larger for smaller $M$, and therefore it takes longer for $\phi$ to converge to $\phi_{\rm min}$. 

\begin{figure}
\epsfxsize=5 in \centerline{\epsfbox{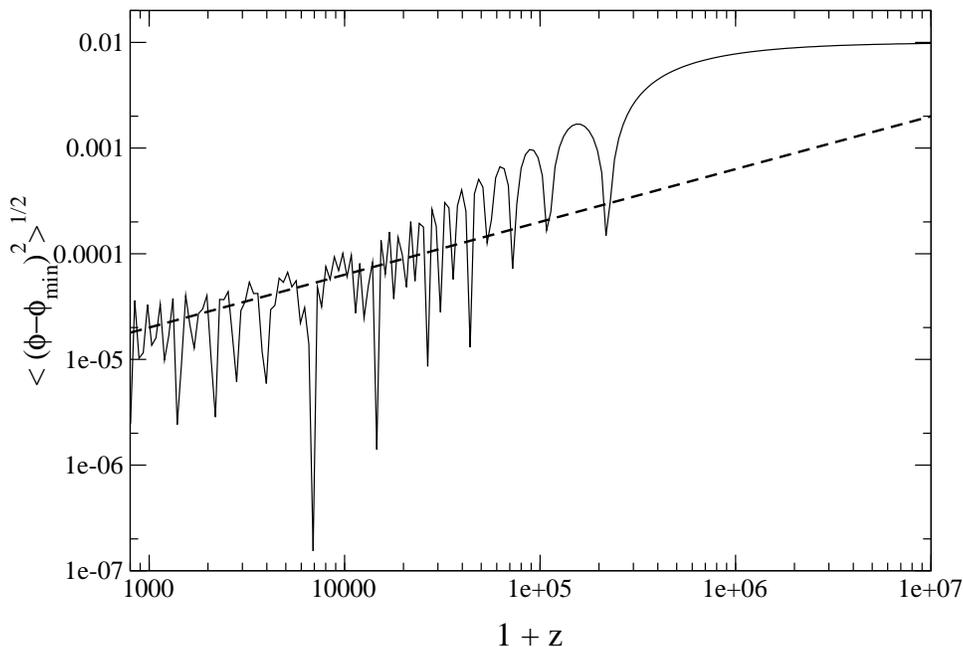}}
\caption{The root mean square quantity $<(\phi-\phi_{\rm min})^2>^{1/2}$ plotted as a function of redshift for the undershoot solution with $\phi_i = 10^{-2}\;M_{\rm Pl}$ and neglecting kicks. The potential is as in the previous two figures. The dotted line is the slope predicted by Eq.~(\ref{tocompare}).}
\label{eg}
\end{figure}

The above results have been checked numerically, as shown in Fig.~\ref{eg}. The dotted line has the slope predicted by Eq.~(\ref{tocompare}). We see that the latter is a good fit throughout the oscillatory regime, even when the oscillations appear to be anharmonic.

Finally, we note in passing that Eqs.~(\ref{rhoavg}), (\ref{mbehave}) and (\ref{tocompare}) together imply
\begin{equation}
\rho_\phi \sim a^{-3(3n+4)/2(n+1)}\,.
\end{equation}
For any $n>0$, we see that the energy density in the oscillations redshifts faster than radiation. Therefore, the chameleon does not suffer from the old moduli problem.

\section{Constraints on initial conditions} \label{constraints}

Next we use the results of the previous section to determine what subset of initial conditions gives a cosmology consistent with current observations. As derived in Sec.~\ref{over} (see Eq.~(\ref{Delphitot})), if the chameleon starts at, or overshoots to $\phi \;\gsim\;\beta M_{\rm Pl}$, then the field will not be at the minimum of 
the effective potential by the beginning of BBN. We now argue that this case is ruled out by BBN constraints on time-variation of particle masses. 

Due to the conformal coupling in Eq.~(\ref{conformal}), a constant mass scale $m^{(i)}$ in the matter-frame is related to a $\phi$-dependent mass scale $m(\phi)$ in Einstein-frame by the rescaling $m(\phi) = e^{\beta_i\phi/M_{\rm Pl}}m^{(i)}$. Thus variations in $\phi$ lead to variations in the various masses:
\begin{equation}
\left\vert\frac{\Delta m}{m}\right\vert \approx \frac{\beta}{M_{\rm Pl}} |\Delta\phi|\,.
\label{deltam}
\end{equation}
Nucleosynthesis constrains the variation in $m$ from the time of nucleosynthesis until today to be less than 10\% or so. Since the value of $\phi$ at the minimum today is much less than $M_{\rm Pl}$, this constrains $\phi_{\rm BBN}$, the value assumed by $\phi$ at BBN, to satisfy
\begin{equation}
\phi_{\rm BBN}\;\lsim\; 0.1 \;\beta^{-1} M_{\rm Pl}\,.
\label{BBN}
\end{equation}
Evidently, if the field starts at, or overshoots to, $\phi \;\gsim \;\beta M_{\rm Pl}$, then it will not yet be at the minimum by the onset of BBN, assuming $\beta\sim {\cal O}(1)$. In this case, $\phi_{\rm BBN}$ violates the above bound. 

Thus, for the undershoot and overshoot cases, we have the respective constraints $\phi_i\;\lsim\;\beta M_{\rm Pl}$ and $\phi_{\rm stop}\;\lsim\;\beta M_{\rm Pl}$. Using Eq.~(\ref{phistop}), the latter translates in a bound on the initial fractional energy density in the chameleon:
\begin{equation}
\Omega_{\phi}^{(i)} \; \lsim \; 1/6\,,
\label{maincons}
\end{equation}
where we have neglected $\phi_i$ and assumed $\beta\sim{\cal O}(1)$.
This is the main result of this Section. 
Remarkably, this is a much weaker bound than in normal quintessence, where $\Omega_{\phi}^{(i)}$ is required by BBN to be less than $10^{-2}$. One would have expected the opposite since, after all, the chameleon couples directly to matter whereas normal quintessence does not. But, as shown above, it is precisely this direct coupling that pushes the field more effectively towards the attractor solution, through a sequence of kicks, thereby resulting in a weaker constraint. The above upper bound on $\Omega_{\phi}^{(i)}$ allows for a range of initial scalar field energy density spanning many orders or magnitude and consistent with equipartition of energy at reheating.

Equation~(\ref{maincons}) ensures that the field is at the minimum from the onset of BBN until the present epoch. In the rest of the Section, we show that it is also a sufficient condition to satisfy all current cosmological constraints and thus to obtain a successful cosmology.

\subsection{Cosmic microwave background anisotropy}\label{cmbsec}

A potentially important effect of the chameleon on the CMB is the modification of the distance to the
last scattering surface. Indeed, Eq.~(\ref{deltam}) implies in particular that the electron mass varies by
\begin{equation}
\left\vert\frac{\Delta m_e}{m_e}\right\vert \approx \frac{\beta}{M_{\rm Pl}} (\phi_{\rm min}^{(0)} - \phi_{\rm min}^{({\rm rec})}) \,,
\label{deltame}
\end{equation}
where $\phi_{\rm min}^{(0)}$ and $\phi_{\rm min}^{({\rm rec})}$ are the field values at the minimum today and at recombination, respectively. Such a variation in the electron mass modifies the binding energy of hydrogen and thus changes the redshift of recombination, $z_{\rm rec}$, by
\begin{equation}
\left\vert\frac{\Delta z_{\rm rec}}{z_{\rm rec}}\right\vert \approx \frac{\beta}{M_{\rm Pl}} (\phi_{\rm min}^{(0)} - \phi_{\rm min}^{({\rm rec})})\,.
\label{deltazrec}
\end{equation}
The WMAP experiment constrains $z_{\rm rec}$ to within 0.1\% or so. 
From Eqs.~(\ref{phimineg}) and~(\ref{phimin0}), we find that $\phi_{\rm min}^{(0)}\gg \phi_{\rm min}^{({\rm rec})}$ and $\phi_{\rm min}^{(0)}/M_{\rm Pl} \sim (M/M_{\rm Pl})^{n/(n+1)}$, which trivially satisfies the WMAP bound. Thus the distance to last scattering is virtually identical to that in normal quintessence.

The other effects of the chameleon on the CMB are the backreaction of its energy density and its influence on the growth of perturbations. The former is trivially satisfied. Indeed, for temperatures less than 10 MeV ($z\;\lsim\;10^{10}$), the energy density in $\phi$ behaves effectively like a (small) cosmological constant, as shown in Section III B. We now turn our attention to the linear evolution of density perturbations. The full non-linear analysis requires numerical work which lies beyond the
scope of the present paper.

\subsection{Density perturbations and large scale structure} \label{lsssec}

We study perturbations in the synchronous gauge, where
the perturbed line element has the form
\begin{equation}
ds^2 = a^2(\tau)\left(-d\tau^2 + (\delta_{ij}+h_{ij}) dx^i dx^j\right)\,.
\end{equation}
Throughout this Section, we use conformal time $\tau$ rather than cosmic time.
The perturbation equations have been discussed in many papers (see, {\it e.g.},~\cite{amendi} and references therein), so we
simply state them here for pressureless dust and the chameleon field. We follow the
literature and write the equations in Einstein frame. 

The evolution equations for the dark matter density contrast, $\delta_c = \delta(\rho_{\rm m}e^{\beta\phi/M_{\rm Pl}})/\rho_{\rm m}e^{\beta\phi/M_{\rm Pl}}$, and the divergence of the velocity field of the dark matter fluid,
$\theta_c = \nabla\cdot \vec{v}$, are respectively given by
\begin{equation} \label{cdmone}
\delta_{c}' = - \left[ \theta_{c} + \frac{h'}{2} \right]
+ \frac{\beta}{M_{\rm Pl}} (\delta \phi)'\,;
\end{equation}
\begin{equation}\label{cdmtwo}
\theta_{c}' = -aH\theta_{c}
+ \beta k^2 \delta\phi - \frac{\beta}{M_{\rm Pl}} \phi' \theta_{c}\,,
\end{equation}
where primes denote derivatives with respect to $\tau$, and $h\equiv \delta^{ij}h_{ij}$ is the trace of $h_{ij}$. For the latter, Einstein's equations give
\begin{equation}\label{eqforh}
h'' + aH h'= -\frac{a^2 \rho_c}{M_{\rm Pl}^2} \delta_c\,.
\end{equation}
Meanwhile, the perturbed Klein--Gordon equation for the chameleon reads
\begin{equation}\label{kgpertur}
(\delta \phi)'' + 2aH(\delta \phi)' +
\left(k^2 + a^2 V_{,\phi\phi} \right)\delta \phi
+ \frac{1}{2}h'\phi' =- \frac{\beta}{M_{\rm Pl}} a^2 \rho_{c}\delta_{c}\,.
\end{equation}
We can safely ignore terms proportional to $\phi'$ since the field is slow-rolling along the attractor, as shown earlier. Then, taking the time-derivative of Eq.~(\ref{cdmone}) and using Eqs.~(\ref{cdmtwo}), (\ref{eqforh}) and~(\ref{kgpertur}), we obtain the following equation for $\delta_c$:
\begin{equation}\label{densitypertur}
\delta_c'' + aH \delta_c' = \frac{3}{2}a^2H^2
\left[ 1 + \frac{2\beta^2}{1+a^2 V_{,\phi\phi}/k^2}\right]\delta_c\,,
\end{equation}
where the quantity in brackets can be interpreted as an effective Newton's constant. In particular, the term proportional to $V_{,\phi\phi}$ results from the chameleon-mediated force~\cite{amendi}, which is negligible if the physical length scale of the perturbation is much larger than the range of the $\phi$-mediated force, that is, if $a/k \gg V_{,\phi\phi}^{-1/2}$. In this case the left hand side of Eq.~(\ref{densitypertur}) is well approximated by $3a^2H^2 \delta_c/2$ and the dark matter fluctuations grow as in GR.

If the field is at the minimum, then $m^2\approx V_{,\phi\phi}\gg H^2$, as shown earlier. For instance, our fiducial potential with $n\geq 1$ gives $m/H > 10^{10}$ at recombination. Thus the chameleon length scale is much less than the Hubble horizon at the time of recombination, which therefore implies that the CMB spectrum is unaffected. This has to be contrasted with normal quintessence models, for which the mass of the field is on the order of the expansion rate along the tracker solution. In this case, fluctuations on scales of the size of the horizon cannot be neglected and leave distinctive features in the anisotropy spectrum of the CMB~ \cite{quintcmb}.

Writing $V(\phi) \approx M^4 + M^{4+n}/\phi^{n}$, the length scale $\lambda_{\rm cham}\equiv V^{-1/2}_{\phi\phi}$ below which perturbations feel a different Newton's constant is given by 
\begin{equation}
\lambda_{\rm cham} =\frac{1}{\sqrt{n(n+1)}} \frac{1}{M} \left(\frac{\phi}{M}\right)^{1 + n/2} \sim 10^{-2} \left(\frac{\phi}{M}\right)^{1+n/2} {\rm cm} \,,
\label{lambdacham}
\end{equation}
where in the last step we have assumed $n\sim{\cal O}(1)$ and substituted $M = 10^{-3}$~eV. 
Density perturbations below that scale will 
initially grow faster because the effective Newton's constant is larger. From Eq.~(\ref{densitypertur}) 
the modified growth rate is $\delta \propto \tau^x$, where 
\begin{equation}
x = -\frac{1}{2} \pm \sqrt{\frac{1}{4} + 6\left(1+2\beta^2\right)}.
\end{equation}
For $\beta = 1$, for example, one obtains  $x = 3.8$ for the growing mode, compared with $x = 2$ 
as predicted by GR. Such perturbations thus grow faster and enter the non--linear 
regime earlier. From Eq.~(\ref{phimin0}) with $M = 10^{-3}$~eV and $n\; \gsim\; 1$, we find that $\lambda_{\rm cham}$ at the present epoch satisfies
\begin{equation}
\lambda_{\rm cham} \;\lsim \; 100\;{\rm pc}\,.
\end{equation}
While this is a rather small length scale, it would be interesting to investigate if this could change substantially the details of galaxy formation, and if early star formation (and therefore early reionization) 
is a natural consequence of chameleon cosmology. Another avenue worthy of investigation is whether the chameleon favors the formation of supermassive black holes at the center of galaxies.

\section{The chameleon during inflation} \label{inflation}

Let us now study the behavior of the chameleon during a period 
of inflation in the early universe. During inflation, the effective potential for the chameleon field is
\begin{equation}
V_{\rm eff}(\phi)\approx  M^4\exp(M^n/\phi^n) + \rho_{\rm vac}e^{4\beta\phi/M_{\rm Pl}}\,,
\end{equation}
where the last term arises from the coupling of the chameleon to the
inflaton vacuum energy, $\rho_{\rm vac}$. The factor of $4$ in the exponential follows from the fact that
$1-3w=4$ in this case. Following the analysis leading to Eq.~(\ref{rhoc}), we find that this potential
has a minimum at some $\phi_{\rm min}\ll M$. Since $\rho_{\rm vac}$ is essentially constant
during inflation, so is $\phi_{\rm min}$, and the chameleon is stabilized.

The mass of the chameleon at the minimum is readily obtained from Eq.~(\ref{mfin}) by letting $\beta\rightarrow 4\beta$ and $\Omega_{\rm m}\rightarrow \Omega_{\rm vac} \approx 1$:
\begin{equation}
m^2 \approx 12\beta n \frac{M_{\rm Pl}}{M}\left(\frac{M}{\phi}\right)^{n+1}H^2\gg H^2\,.
\label{mfininf}
\end{equation}
As the mass is constant during inflation, the chameleon oscillates
around the minimum with an average amplitude given by
\begin{equation}
<(\phi -\phi_{\rm min})^2>\approx a^{-3}\,.
\end{equation}
As expected, the chameleon field behaves like a dust component during inflation. Due to the exponential growth of the scale factor, this implies that $\phi$ very quickly settles at the minimum of the potential. 

At the end of inflation when the universe reheats, the inflaton decays not only into radiation and matter, but also into coherent kinetic energy of the chameleon. However, the production of chameleon quanta is generally suppressed
because of their large mass. Since the universe becomes radiation-dominated at the end of inflation, the minimum
of the effective potential suddenly moves to a much larger value. The chameleon is therefore released 
from a field value much smaller than $\phi_{\rm min}$. Hence, we conclude that the overshoot solution, discussed in Section~\ref{overshoot}, is the more natural outcome of an inflationary phase.

\section{Chameleon and time-varying $\alpha$} \label{alpha}

Recent analysis of absorption spectra of quasars have led some to claim that the fine-structure constant $\alpha_{\rm EM}$ might have evolved by approximately one part in $10^5$ over the redshift range $0.2<z<3.7$~\cite{webb}. If this turns out to be true, then general covariance would imply that $\alpha_{\rm EM}$ can vary both in space and in time, that is, it must be a function of a field. A space-time varying fine-structure constant was first posited by Jordan~\cite{jordan}, Teller~\cite{teller}, Stanyukovich~\cite{stan}, while the implications for EP violations were studied by Dicke~\cite{dicke} and more recently by Bekenstein~\cite{beken}. The recent evidence from quasar spectra has rejuvenated this idea, resulting in a flurry of papers~\cite{all,uzan}.

Here we show that the chameleon cannot lead to a time-variation of $\alpha_{\rm EM}$ consistent with the recent observational claims. Since we have so far assumed that the chameleon couples conformally to matter fields, as seen from Eq.~(\ref{conformal}), and since the Maxwell action is conformally-invariant, at tree-level the chameleon does not lead to a time-varying $\alpha_{\rm EM}$. However, one can easily consider generalizations to the matter action in Eq.~(\ref{action}), such as
\begin{equation}
\int d^4x e^{\tilde{\beta}\phi/M_{\rm Pl}}{\cal L}_{\rm m}(\psi_{\rm m}^{(i)},g_{\mu\nu}^{(i)})\,,
\label{mattermod}
\end{equation}
where $\tilde{\beta}$ is a constant. In particular, the Maxwell lagrangian in this case reads $g^{-2}e^{\tilde{\beta}\phi/M_{\rm Pl}}F_{\mu\nu}F^{\mu\nu}$, corresponding to
\begin{equation}
\alpha_{\rm EM}\sim e^{-\tilde{\beta}\phi/M_{\rm Pl}}\,.
\end{equation}

Then, variations in $\phi$ induce variations in $\alpha_{\rm EM}$ of magnitude
\begin{equation}
\left\vert\frac{\Delta \alpha_{\rm EM}}{\alpha_{\rm EM}}\right\vert \approx \tilde{\beta} \frac{|\Delta\phi|}{M_{\rm Pl}}\,.
\label{deltanu}
\end{equation}
Since the cosmological value of the chameleon, $\phi(z)$, is a decreasing function of density and, therefore, a decreasing function of redshift $z$, we can approximate Eq.~(\ref{deltanu}) by
\begin{equation}
\left\vert\frac{\Delta \alpha_{\rm EM}}{\alpha_{\rm EM}}\right\vert \approx \tilde{\beta}\frac{\phi^{(0)}_{\rm min}}{M_{\rm Pl}}\sim \tilde{\beta}\left(\frac{M}{M_{\rm Pl}}\right)^{n/(n+1)}\,,
\label{deltanu2}
\end{equation}
where we used Eq.~(\ref{phimin0}) in the last step. For $n\;\gsim\; 1$, the resulting $\Delta\alpha_{\rm EM}/\alpha_{\rm EM}$ is many orders of magnitude too small to account for the time-variation advocated by~\cite{webb}.

Taking into account local variations in the chameleon does not help. Indeed, recall that for sufficiently dense objects, the local value of the chameleon depends mostly on the local density and is insensitive to the background cosmic density. The absorption clouds used in the analysis of~\cite{webb} can be classified in three populations: weak MgII systems, Lyman limit systems and damped Lyman-alpha systems, with estimated density of $10^{-25}$, $10^{-26}$ and $10^{-24}-10^{-23}$~g/cm$^3$, respectively. The key point is that all of these are comparable with the local galactic density of $10^{-24}$~g/cm$^3$, and thus the chameleon value in these systems should be nearly the same as locally.

\section{Application to the Cyclic Universe Model} \label{cyclicpot}

To illustrate the usefulness of the chameleon mechanism in designing cosmological scenarios, in this Section we apply our results to the cyclic model of the universe~\cite{cyclic,ek,seiberg,pert,design,rest}. The cyclic scenario proposes that time did not begin at the big bang, as assumed implicitly in standard inflationary cosmology, but rather extends infinitely far in the past as well as in the future. Thus our current epoch of expansion is only one out of an infinite number of cycles. The cyclic model addresses the homogeneity, flatness and monopole problems of the big bang model, and generates a nearly scale-invariant spectrum of density fluctuations, without invoking a period of high-energy accelerated expansion. As such, it constitutes the most serious candidate for a viable alternative to the inflationary paradigm.

Except for very near the big crunch/bang transition, the cyclic model is well-described by a four-dimensional effective action of the form given in Eq.~(\ref{action}):
\begin{equation}
S=\int d^4x\sqrt{-g}\left\{\frac{M_{\rm Pl}^2}{2}{\cal
R}-\frac{1}{2}(\partial\phi)^2- V(\phi)\right\}
- \int d^4x{\cal L}_{\rm m}(\psi_{\rm m}^{(i)},g_{\mu\nu}^{(i)})\,.
\label{actioncyclic}
\end{equation}
In particular, as in the simplest inflationary models, the scenario consists of a scalar field $\phi$ rolling down its potential $V(\phi)$. An important distinction with inflationary theory, however, is that here $\phi$ has a higher-dimensional interpretation of measuring the distance $d$ between two end-of-the-world branes (more precisely, orbifold fixed planes). In other words, $\phi$ is a radion. In this higher-dimensional picture, the big crunch/bang bridge between each cycle corresponds to the cataclysmic collision of the branes. The relation is $d = L\ln[\coth(\phi/M_{\rm Pl}\sqrt{6})]$~\cite{ek,cyclic}, and thus the brane collision ($d\rightarrow 0$) corresponds to the limit $\phi\rightarrow \infty$.

At tree-level, the metrics $g_{\mu\nu}^{(i)}$ are given by Eq.~(\ref{conformal}) with couplings
\begin{equation}
\beta_i = \frac{1}{\sqrt{6}}\,,
\label{betacyclic}
\end{equation}
which corresponds to the Kaluza-Klein limit in the higher-dimensional picture. We henceforth assume that this regime holds for the relevant range of $\phi$.
The potential $V(\phi)$ is thought to arise from non-perturbative effects, such as virtual exchange of branes in the higher-dimensional theory. A typical cyclic potential, sketched in Fig.~\ref{cyclic}, is given by
\begin{equation}
V(\phi) = M^4e^{M^n/(\phi-\phi_\star)^n}\left\{1 - e^{-c(\phi-\phi_{\rm cross})/M_{\rm Pl}}\right\}\cdot {\cal F}(\phi)\,,
\label{cycfidpot}
\end{equation}
where we have reintroduced $\phi_\star$, the value of $\phi$ for which the potential diverges. The field value $\phi_{\rm cross}$ is where the potential vanishes. The positive constant $c$ must satisfy $c\;\gsim\; {\cal O}(10)$ in order for the spectrum of density perturbations to be nearly scale invariant~\cite{pert}. The function ${\cal F}(\phi)$ accounts for the fact that the non-perturbative effects must turn off as $\phi\rightarrow\infty$, since the string coupling goes to zero in this limit. In order for the Kaluza-Klein limit assumed in Eq.~(\ref{betacyclic}) to apply for all $\phi>\phi_\star$, we impose that
\begin{equation}
\phi_\star\gg M_{\rm Pl}\,.
\end{equation}

Currently, the field lies at $\phi^{(0)}-\phi_\star\gg M$, indicated by a dot in the Figure.
Thus, $V(\phi_0)\approx M^4$, and this potential energy drives the observed acceleration of the universe (Region a)). This phase of cosmic acceleration lasts sufficiently long to empty out our observable universe, thereby making it highly homogeneous, isotropic, spatially-flat and nearly vacuous. After a while, the field begins to roll down the potential and reaches Region b). Since $V<0$ in this region, cosmic expansion eventually comes to halt, and the universe enters a phase of contraction. It is during this phase that a nearly scale-invariant spectrum of density perturbations is generated from quantum fluctuations in $\phi$~\cite{ek,pert}. When the field reaches Region c), the function ${\cal F}(\phi)$ in Eq.~(\ref{cycfidpot}) becomes important and causes $V$ to go to zero~\cite{note}. This ensures that the energy density of the universe is dominated by the kinetic energy in $\phi$ as it zooms toward $+\infty$, as required by the prescription of~\cite{seiberg}. The $\phi \rightarrow\infty$ limit corresponds to the big crunch/bang transition, at which point the universe reheats and becomes filled with thermal matter and radiation. This marks the beginning of the hot big bang phase. Meanwhile, $\phi$ bounces back, rushes through Region b) and eventually comes to a stop in Region a). It remains essentially frozen there, until the universe is sufficiently cold to allow the vacuum energy in $\phi$ to drive cosmic acceleration. The cycle then repeats itself.

It is crucial that $\phi$ does not result in EP violations stronger than allowed by experiments. One way to ensure this, as proposed by Steinhardt and Turok~\cite{cyclic}, is if the couplings $\beta_i$ in Eq.~(\ref{betacyclic}) are functions of $\phi$ and satisfy
\begin{equation}
\beta_i(\phi^{(0)}) \; \lsim \; 10^{-4}\,,
\label{betacond}
\end{equation}
for today's value of the field, $\phi^{(0)}$. This possibility is certainly allowed by some brane-world models, such as the Randall-Sundrum scenario~\cite{rs}. In general, however, one expects that $\beta_i$ will be of order unity and different for different matter species.

Here we argue that no such extra condition is necessary since $\phi$ is in fact a chameleon. Indeed, the reasoning that lead to Eq.~(\ref{betacond}) neglected the effect of the background matter density. In Region a), the potential is approximately given by
\begin{equation}
V(\phi) \approx  M^4e^{M^n/(\phi-\phi_\star)^n}\,,
\end{equation}
which is of the same form as our fiducial potential (see Eq.~(\ref{fidpot})). Moreover, since the various $\beta_i$'s are non-zero and positive, the dynamics of $\phi$ are governed by the effective potential in Eq.~(\ref{veff}) with $\beta_i=1/\sqrt{6}$.

The above analysis therefore greatly expands the range of models and brane-world set-ups suitable for cyclic cosmology. The only requirements are that $\beta_i>0$ and that the potential be of the general form
\begin{equation}
V(\phi) = M^4f(\phi/M)\left\{1 - e^{-c(\phi-\phi_{\rm cross})/M_{\rm Pl}}\right\}\cdot {\cal F}(\phi)\,,
\label{cycfidpot2}
\end{equation}
where $f$ satisfies Eqs.~(\ref{condsf}). In particular, the chameleon mechanism disposes of condition~(\ref{betacond}).

\section{Conclusions}

We have explored the complete cosmological evolution of the chameleon field for general
models where the effective potential displays a minimum, as in the original scenario.
Our analysis shows that the chameleon can act as a dark energy particle at late times, accounting for the current phase of cosmic acceleration.
We have found that the minimum is an attractor with undershoot and overshoot 
solutions, analogous to normal quintessence models. In studying the approach to the
attractor, it is important to take into account the kicks due to species
becoming non-relativistic during the radiation era. These kicks successively push the
chameleon field towards the minimum of the potential and consequently greatly expand the
basin of attraction. It would be interesting to study the role played by
the kicks in models of quintessence coupled to dark matter.

The most stringent constraint on our model comes from time variation of masses since BBN. 
This requires the chameleon to have settled to the minimum by the onset of BBN.
If this is not realized, then the electron kick will result in
an unacceptably large variation in masses. This condition is fulfilled for a broad range of initial conditions spanning many orders of magnitude in initial chameleon energy density. The allowed range is in fact broader than in normal quintessence, largely due to the kicks which make the attractor mechanism comparatively more efficient.

We have studied the chameleon during inflation and find that it quickly stabilizes at the minimum as the universe inflates. We have argued that the chameleon cannot lead to a time variation of the fine-structure constant sufficiently large to be consistent with recent observational claims. Finally we applied the chameleon to the cyclic universe. When the chameleon
mechanism is taken into account the class of potentials relevant to the
cyclic universe is enlarged and constraints on the parameters relaxed.

\section{Acknowledgements}

We thank L.~Amendola, C.~Churchill, P.-S.~Corasaniti, D.~Mota, J.~Murugan, C.S.~Rhodes, C.~Skordis, P.J.~Steinhardt and D.~Tocchini-Valentini for insightful discussions. A.W. is grateful to the KITP at UC Santa Barbara and the organizers of the String Cosmology program for their hospitality. This work was supported in part by PPARC (CvdB and ACD), the British Council Alliance exchange grant (PB, CvdB and ACD), the Columbia University Academic Quality Fund, the Ohrstrom Foundation (JK), DOE grant DE-FG02-92ER40699, the Pfister Foundation, and the University of Cape Town (AW).

\section{Appendix} \label{append}

\subsection{The Thin-Shell property} \label{apa}

We have stated in Sec.~\ref{Mrev} that $M\le 10^{-3}$~eV
results from local tests of gravity. We provide more details
here on how this bound is obtained. Consider a spherical body of homogeneous density $\rho_c$, radius $R_c$ and total mass $M_c=4\pi R_c^3\rho_c/3$, immersed in a homogeneous medium of density $\rho_\infty$. We denote by $\phi_c$ and $\phi_\infty$ the field values which minimize the effective potential for the respective densities. At short distances, the total force $F$, gravitational plus chameleon-mediated, on a test mass is~\cite{cham}
\begin{equation}
F= (1+\theta) F_{\rm N}\,,
\end{equation}
where $F_{\rm N}$ is Newtonian force and $\theta$ is the fractional force due to the chameleon. For small objects, in a way to be made precise below, $\theta=2\beta^2$, the usual answer for a scalar field without potential. For much larger objects, however, one finds that
\begin{equation}
\theta = 2\beta^2\frac{\phi_\infty-\phi_c}{2\beta M_{\rm Pl}\Phi_c}\,,
\end{equation}
where $\Phi_c= M_c/8\pi M_{\rm Pl}^2R_c$ is the Newtonian potential of the object.
Thus, for sufficiently large objects such that $(\phi_\infty-\phi_c)/(2\beta M_{\rm Pl}\Phi_c) \ll 1$, one has
$\theta\ll 2\beta^2$ and the fifth force is suppressed. If $(\phi_\infty-\phi_c)/(2\beta M_{\rm Pl}\Phi_c) \gg 1$, on the other hand, one simply gets $\theta=2\beta^2$.

The suppression for large bodies is due to the so-called thin shell effect which can be understood as follows. 
Essentially, only a thin shell under the surface of the object contributes to the $\phi$ pull on a test mass.
Indeed, one finds that the profile of the chameleon field is such that it is nearly constant up to a radius
$R_s< R_c$, where $R_s$ is given by
\begin{equation}
\frac{R_c -R_s}{R_c}= \frac{\phi_\infty-\phi_c}{6\beta M_{\rm Pl}\Phi_c}\,.
\end{equation}
Whether an object has a thin shell or not, that is whether its chameleon-mediated force is suppressed compared to the usual answer, depends on the magnitude of this ratio. The shell is thin if
\begin{equation}
\frac{R_c -R_s}{R_c}\ll 1\,.
\end{equation}
Since $\phi_c\ll\phi_\infty$ for large density contrast between the object and the ambient matter, this generally reduces to
\begin{equation}
\frac{\phi_\infty}{M_{\rm Pl}} \ll \Phi_c \,.
\end{equation}
For the fiducial potential $V(\phi) = M^4\exp(M^n/\phi^n)$ with $n\sim {\cal O}(1)$, applying this condition for typical test masses used in laboratory tests of gravity leads to the constraint
\begin{equation} 
M\;\lsim \; 10^{-3}\;{\rm eV}\,.
\end{equation}

\subsection{The Instantaneous Kick} \label{apkick}
 
In this Section, we present an analytic approximation to the contribution from each mass threshold to $T^\mu_\mu$ during the radiation dominated era, as described in Sec.~\ref{undershoot}.
The idea is to approximate each contribution as a delta-function source, that is, as an instantaneous kick. Specifically, the equation of motion for $\phi$ is approximated by
\begin{equation} 
\ddot \phi + 3H \dot \phi + V_{,\phi} = -\beta \kappa H M_{\rm Pl} \delta (t-t_0) 
\label{eomdelta}
\end{equation} 
where the right hand side corresponds to a species contributing sharply around time $t_0$. 
The constant $\kappa$ is of order $g_i/g_\star(m_i)$, in the notation of Sec.~\ref{undershoot}.
The order of magnitude of the delta function was estimated using the fact that the 
contribution to $T^\mu_\mu$ is of order $H^2$ ({\it i.e.}, the $\tau$ function peaks at a value of order unity) and that the width of the $\tau$ function is of order of a fraction of a Hubble time.

First, let us consider the regime where $\phi$ is much larger than $\phi_{\rm min}$. 
In this case, Eq.~(\ref{eomdelta}) becomes 
\begin{equation} 
\ddot \phi + 3H \dot \phi =- \beta\kappa H M_{\rm Pl} \delta (t-t_0) \,,
\end{equation} 
as the potential is negligible. 
Assuming the field is initially at rest, the solution is 
\begin{equation} 
t<t_0:\qquad \phi = \phi_i\,,
\end{equation} 
and 
\begin{equation} 
t\ge t_0:\qquad \phi= (1 - a_0^3 \beta \kappa H_0 M_{\rm Pl} )\int_{t_{0}}^{t} a^{-3}dt + \phi_i \,,
\end{equation} 
where $a_0$ and $H_0$ are evaluated at $t_0$. 
The kick thus results in a jump in $\phi$ from its free evolution of
\begin{equation} 
\Delta\phi= -a_0^3 \beta\kappa H_0 M_{\rm Pl} \int_{t_{0}}^{t} a^{-3}dt \,,
\end{equation} 
which, in the radiation-dominated era, converges to
\begin{equation} 
\Delta\phi = -\kappa \beta M_{\rm Pl} \,.
\end{equation} 
Recalling that $\kappa\sim g_i/g_\star(m_i)$, this agrees with Eq.~(\ref{Deli}).

\begin{figure}
\epsfxsize=5 in \epsfysize=5 in \centerline{\epsfbox{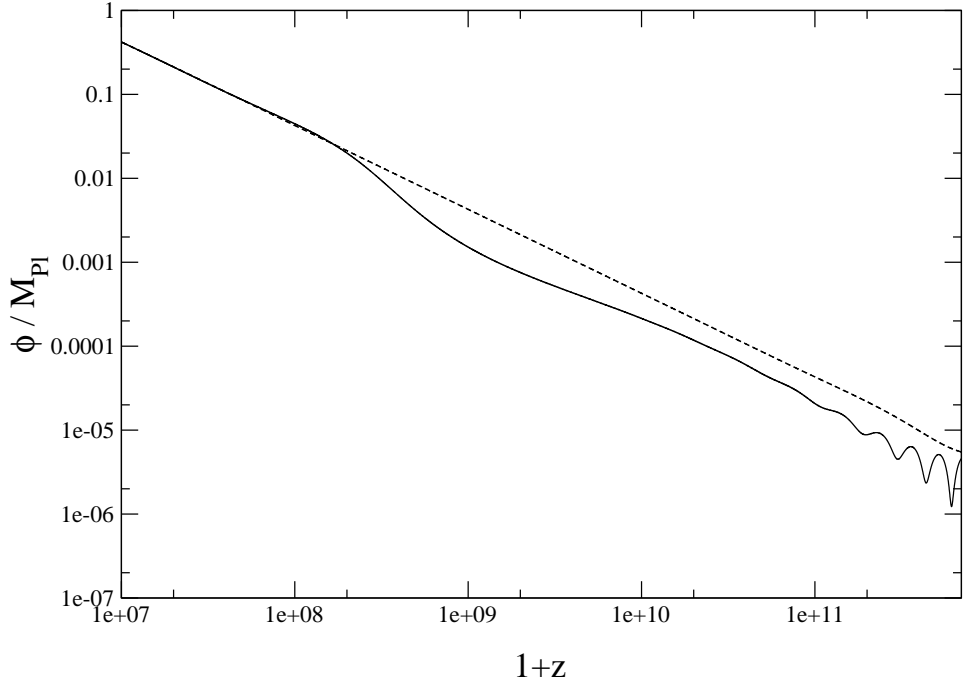}}
\caption{The chameleon starts near the minimum and is subjected to the contribution to $T^\mu_\mu$ from some particle species becoming non-relativistic. In this example, the potential is $V(\phi) = M^6/\phi^2$, $\beta = 1$ and $M = 10$ MeV, the latter once again chosen to be much larger than the required $M=10^{-3}$~eV due to numerical limitations. We see that the field oscillates and eventually settles back to the minimum.}
\label{ekick}
\end{figure}

Next, consider the effect of the kick if the field is close to or at the minimum.
In this case, we can linearize the effective potential about the minimum and obtain 
\begin{equation} 
\ddot \phi + 3H \dot \phi +m^2(\phi-\phi_{\rm min})=- \beta\kappa H M_{\rm Pl} \delta (t-t_0)\,. 
\end{equation} 
Letting $\delta \phi \equiv \phi -\phi_{\rm min}= a^{-3/2}\psi$, we find 
\begin{equation} 
\ddot \psi + \left(m^2 +\frac{3H^2}{4}\right) \psi = - \beta\kappa H M_{\rm Pl} a^{3/2} \delta (t-t_0) \,.
\end{equation} 
Assuming the field is at the minimum initially, $\psi=0$, and using $m^2 \gg H^2$, the solution reads 
\begin{equation} 
t< t_0:\qquad \psi= 0\,, 
\end{equation} 
and 
\begin{equation} 
t\ge t_0: \qquad \psi = - \frac{\beta\kappa H_0 M_{\rm Pl}}{m} \sin (m(t-t_0)) \,.
\end{equation} 
Averaging over many oscillations, we find 
\begin{equation} 
<(\phi-\phi_{\rm min})^2> = \frac{\beta^2\kappa^2 H_0^2 M_{\rm Pl}^2}{2m^2}\left(\frac{a_0}{a}\right)^3\,.
\end{equation} 
Thus the kick makes the chameleon oscillate about the minimum. These oscillations are 
damped due to the expansion of the universe, and the field quickly settles back to the minimum.
This is confirmed by solving the equations of motion numerically, using the exact form for the $\tau$ function, as shown in Fig.~\ref{ekick}.

Notice that the amplitude of the oscillations is always small compared to $M_{\rm Pl}$.
In particular, using Eq.~(\ref{deltam}), the average variation of masses due to the kick behaves like 
\begin{equation} 
\left<\left(\frac{\Delta m}{m}\right)^2\right>= \frac{\beta^4\kappa^2 H_0^2 }{2m^2}\left(\frac{a_0}{a}\right)^3\ll 1\,, 
\end{equation} 
implying no significant modification to BBN due to the electron kick, for instance. 
 
\subsection{Convergence to the Minimum} \label{apb}

Here we wish to prove Eq.~(\ref{interm}) directly. If we assume that the scale factor is power-law in time, $a(t) \sim t^p$, then Eq.~(\ref{mbehave}) implies that the mass of the chameleon can be written as
\begin{equation}\label{moft}
m(t) = \frac{B}{t^q}\,,
\end{equation}
where $B$ and $q$ are positive constants. 

Small fluctuations $\phi_1(t) = \phi(t) - \phi_{\rm min}(t)$ around the minimum $\phi_{\rm min}$ are then governed by
\begin{equation}
\ddot \phi_1 + \frac{3p}{t} \dot \phi_1 + \frac{B^2}{t^{2q}} \phi_1 = 0\,,
\end{equation}
with general solution
\begin{equation}
\phi_1(t) \sim t^{(1-3p)/2} Z_r\left(\frac{B}{1-q} t^{1-q} \right),
\end{equation}
where $Z_r$ is a superposition of Bessel functions of order $r \equiv (1-3p)/2(1-q)$. Without loss of generality we take $Z_r = J_r$.
Note that the argument of the Bessel function is large since $m(t)t/(1-q)\sim m/H\gg 1$.
Thus, using the asymptotics of the Bessel function, we find
\begin{equation}
\phi_1 (t)  \sim t^{(1-3p)/2} \sqrt{\frac{2(1-q)}{\pi B t^{1-q}}}
\cos\left(\frac{m(t)t}{1-q}\right)\,,
\end{equation}
where we have dropped an irrelevant phase factor.

Averaging $\phi_1^2$ over a few oscillations, while noting that
$m(t)$ is nearly constant during that time (since $\dot{m}/m\sim H\ll m$, as seen from Eq.~(\ref{mdot})),
we get
\begin{equation}
<\phi_1^2 (t)> \sim t^{-3p + q}\,.
\end{equation}
Using Eq.~(\ref{moft}), we obtain the desired result:
\begin{equation}
m(t) <\phi_1^2> \propto t^{-3p} \propto a^{-3}\,,
\end{equation}
in agreement with Eq.~(\ref{interm}). It is straightforward to prove that this also holds for the case $q=1$.

\end{document}

%% file: myepsf.tex
\ifx\epsfannounce\undefined \def\epsfannounce{\immediate\write16}\fi
 \epsfannounce{This is `epsf.tex' v2.7k <10 July 1997>}%
\newread\epsffilein    
\newif\ifepsfatend     
\newif\ifepsfbbfound   
\newif\ifepsfdraft     
\newif\ifepsffileok    
\newif\ifepsfframe     
\newif\ifepsfshow      
\epsfshowtrue          
\newif\ifepsfshowfilename 
\newif\ifepsfverbose   
\newdimen\epsfframemargin 
\newdimen\epsfframethickness 
\newdimen\epsfrsize    
\newdimen\epsftmp      
\newdimen\epsftsize    
\newdimen\epsfxsize    
\newdimen\epsfysize    
\newdimen\pspoints     
\def\epsfbox#1{\global\def\epsfllx{72}\global\def\epsflly{72}%
   \global\def\epsfurx{540}\global\def\epsfury{720}%
   \def\lbracket{[}\def\testit{#1}\ifx\testit\lbracket
   \let\next=\epsfgetlitbb\else\let\next=\epsfnormal\fi\next{#1}}%
\def\epsfgetlitbb#1#2 #3 #4 #5]#6{%
   \epsfgrab #2 #3 #4 #5 .\\%
   \epsfsetsize
   \epsfstatus{#6}%
   \epsfsetgraph{#6}%
}%
\def\epsfnormal#1{%
    \epsfgetbb{#1}%
    \epsfsetgraph{#1}%
}%
\newhelp\epsfnoopenhelp{The PostScript image file must be findable by
TeX, i.e., somewhere in the TEXINPUTS (or equivalent) path.}%
\def\epsfgetbb#1{%
    \openin\epsffilein=#1
    \ifeof\epsffilein
        \errhelp = \epsfnoopenhelp
        \errmessage{Could not open file #1, ignoring it}%
    \else                       
        {
            \chardef\other=12
            \def\do##1{\catcode`##1=\other}%
            \dospecials
            \catcode`\ =10
            \epsffileoktrue         
            \epsfatendfalse     
            \loop               
                \read\epsffilein to \epsffileline
                \ifeof\epsffilein 
                \epsffileokfalse 
            \else                
                \expandafter\epsfaux\epsffileline:. \\%
            \fi
            \ifepsffileok
            \repeat
            \ifepsfbbfound
            \else
                \ifepsfverbose
                    \immediate\write16{No BoundingBox comment found in %
                                    file #1; using defaults}%
                \fi
            \fi
        }
        \closein\epsffilein
    \fi                         
    \epsfsetsize                
    \epsfstatus{#1}%
}%
\def\epsfclipoff{\def\epsfclipstring{\ifepsfdraft\space clip\fi}}%
\epsfclipoff 
\def\epsfspecial#1{%
     \epsftmp=10\epsfxsize
     \divide\epsftmp\pspoints
     \ifnum\epsfrsize=0\relax
       \includegraphics{\ifepsfdraft}%
     \else
       \epsfrsize=10\epsfysize
       \divide\epsfrsize\pspoints
       \includegraphics{\ifepsfdraft}%
     \fi
}%
\def\epsfframe#1%
{%
  \leavevmode                   
  \setbox0 = \hbox{#1}%
  \dimen0 = \wd0                                
  \advance \dimen0 by 2\epsfframemargin         
  \advance \dimen0 by 2\epsfframethickness      
  \vbox
  {%
    \hrule height \epsfframethickness depth 0pt
    \hbox to \dimen0
    {%
      \hss
      \vrule width \epsfframethickness
      \kern \epsfframemargin
      \vbox {\kern \epsfframemargin \box0 \kern \epsfframemargin }%
      \kern \epsfframemargin
      \vrule width \epsfframethickness
      \hss
    }
    \hrule height 0pt depth \epsfframethickness
  }
}%
\def\epsfsetgraph#1%
{%
   \leavevmode
   \hbox{
     \ifepsfframe\expandafter\epsfframe\fi
     {\vbox to\epsfysize
     {%
        \ifepsfshow
            \vfil
            \hbox to \epsfxsize{\epsfspecial{#1}\hfil}%
        \else
            \vfil
            \hbox to\epsfxsize{%
               \hss
               \ifepsfshowfilename
               {%
                  \epsfframemargin=3pt 
                  \epsfframe{{\tt #1}}%
               }%
               \fi
               \hss
            }%
            \vfil
        \fi
     }%
   }}%
   \global\epsfxsize=0pt
   \global\epsfysize=0pt
}%
\def\epsfsetsize
{%
   \epsfrsize=\epsfury\pspoints
   \advance\epsfrsize by-\epsflly\pspoints
   \epsftsize=\epsfurx\pspoints
   \advance\epsftsize by-\epsfllx\pspoints
   \epsfxsize=\epsfsize{\epsftsize}{\epsfrsize}%
   \ifnum \epsfxsize=0
      \ifnum \epsfysize=0
        \epsfxsize=\epsftsize
        \epsfysize=\epsfrsize
        \epsfrsize=0pt
      \else
        \epsftmp=\epsftsize \divide\epsftmp\epsfrsize
        \epsfxsize=\epsfysize \multiply\epsfxsize\epsftmp
        \multiply\epsftmp\epsfrsize \advance\epsftsize-\epsftmp
        \epsftmp=\epsfysize
        \loop \advance\epsftsize\epsftsize \divide\epsftmp 2
        \ifnum \epsftmp>0
           \ifnum \epsftsize<\epsfrsize
           \else
              \advance\epsftsize-\epsfrsize \advance\epsfxsize\epsftmp
           \fi
        \repeat
        \epsfrsize=0pt
      \fi
   \else
     \ifnum \epsfysize=0
       \epsftmp=\epsfrsize \divide\epsftmp\epsftsize
       \epsfysize=\epsfxsize \multiply\epsfysize\epsftmp
       \multiply\epsftmp\epsftsize \advance\epsfrsize-\epsftmp
       \epsftmp=\epsfxsize
       \loop \advance\epsfrsize\epsfrsize \divide\epsftmp 2
       \ifnum \epsftmp>0
          \ifnum \epsfrsize<\epsftsize
          \else
             \advance\epsfrsize-\epsftsize \advance\epsfysize\epsftmp
          \fi
       \repeat
       \epsfrsize=0pt
     \else
       \epsfrsize=\epsfysize
     \fi
   \fi
}%
\def\epsfstatus#1{
   \ifepsfverbose
     \immediate\write16{#1: BoundingBox:
                  llx = \epsfllx\space lly = \epsflly\space
                  urx = \epsfurx\space ury = \epsfury\space}%
     \immediate\write16{#1: scaled width = \the\epsfxsize\space
                  scaled height = \the\epsfysize}%
   \fi
}%
{\catcode`\%=12 \global\let\epsfpercent=
\global\def\epsfatend{(atend)}%
\long\def\epsfaux#1#2:#3\\%
{%
   \def\testit{#2}
   \ifx#1\epsfpercent           
       \ifx\testit\epsfbblit    
            \epsfgrab #3 . . . \\%
            \ifx\epsfllx\epsfatend 
                \global\epsfatendtrue
            \else               
                \ifepsfatend    
                \else           
                    \epsffileokfalse
                \fi
                \global\epsfbbfoundtrue
            \fi
       \fi
   \fi
}%
\def\epsfempty{}%
\def\epsfgrab #1 #2 #3 #4 #5\\{%
   \global\def\epsfllx{#1}\ifx\epsfllx\epsfempty
      \epsfgrab #2 #3 #4 #5 .\\\else
   \global\def\epsflly{#2}%
   \global\def\epsfurx{#3}\global\def\epsfury{#4}\fi
}%
\def\epsfsize#1#2{\epsfxsize}%

%% file: cosmo12.bbl
\begin{thebibliography}{hello}
\bibitem{note}
The prescription of~\cite{seiberg} also holds if the potential remains negative and even diverges in the limit $\phi\rightarrow\infty$, as shown explicitly in~\cite{design}.
\end{thebibliography}
